\documentclass[a4paper,12pt]{elsarticle}
\usepackage{graphicx}
\usepackage{rotating}
\usepackage{booktabs}
\usepackage{romannum}
\usepackage{float}
\usepackage{array}
\usepackage{amsmath}
\usepackage{amsthm}
\usepackage{amsfonts}
\usepackage{amssymb}         
\usepackage{tikz}
\usepackage[most]{tcolorbox}
\usetikzlibrary{intersections}
\usepackage{psfrag}
\usepackage{epstopdf}
\usepackage{caption}
\usepackage{enumerate}    
\usepackage{enumitem}
\usepackage{xcolor, colortbl}  
\usepackage{array}
\usepackage{makecell}
\usepackage{multirow}
\usepackage{tikz}
\usetikzlibrary{calc,fit,shapes,arrows,positioning,chains,arrows.meta}
\usepackage{makeshape}
\usepackage{floatflt}
\usepackage{lscape}
\usepackage{rotating}
\usepackage{multicol}
\usepackage{multirow,bigdelim}
\usepackage{xcolor}
\usepackage{threeparttable}
\usepackage{float}

\usepackage[redeflists]{IEEEtrantools}
\usepackage{hyperref}  
\usepackage{lscape}  
\hypersetup{
	colorlinks = true,
	linkcolor = blue,
	anchorcolor = blue,
	citecolor = red,
	filecolor = blue,
	menucolor = blue,
	runcolor = blue,
	urlcolor = blue
}

\DeclareGraphicsExtensions{jpg,pdf,eps}

\usepackage{geometry}  
\newcommand{\eval}[2][\right]{\relax\ifx#1\right\relax \left.\fi#2#1\rvert}

\makeatletter
\newcommand{\Rmnum}[1]{\expandafter\@slowromancap\romannumeral #1@}
\makeatother

\usepackage[utf8]{inputenc}
\usepackage{xcolor}  
\usepackage{epstopdf}
\usepackage{caption}
\usepackage{subcaption}
\usepackage{natbib}
\usepackage{capt-of}
\usepackage{upgreek}
\usepackage{orcidlink}
\biboptions{sort&compress}
\makeatletter
\def\ps@pprintTitle{%
   \let\@oddfoot\@empty
   \let\@evenfoot\@empty
}
\makeatother
\begin{document}
	\begin{frontmatter}
		
		
        \title{Second-Order Homogenization of Particulate Elastomeric Composites for Structural-Scale Damage Analysis}
		
		\author[1]{Krishna Murthy Pabbu\orcidlink{/0009-0000-4289-8873}}
        \author[1]{Tarkes Dora Pallicity\corref{cor1}\orcidlink{/0000-0003-4445-0305}}
	    \ead{tarkes.pallicity@iitg.ac.in}
	    \author[1,2]{Nelson Muthu\orcidlink{/0000-0002-2589-3234}}
	    \cortext[cor1]{Corresponding Author}
		\affiliation[1]{organization={Indian Institute of Technology Guwahati},
			addressline={Department of Mechanical Engineering}, 
			city={Guwahati},
			postcode={781039}, 
			state={Assam},
			country={India}}
		\affiliation[2]{organization={Indian Institute of Technology Palakkad},
			addressline={Department of Mechanical Engineering}, 
			city={Palakkad},
			postcode={678623}, 
			state={Kerala},
			country={India}}
		\begin{abstract}
				Damage prediction in particulate elastomeric composites is challenging due to the strong material heterogeneity, nearly incompressible behavior of the matrix, and the presence of localized stress concentrations arising from matrix-inclusion interactions. In the present work, a computational framework is developed for damage propagation analysis of particulate elastomeric composite structures. The effective constitutive response of the heterogeneous material is first determined using the second-order homogenization method. The resulting homogenized response is subsequently used to identify equivalent hyperelastic material parameters for structural simulations.  A three-parameter isotropic damage model is incorporated to capture damage-induced softening. A mixed displacement-pressure finite element formulation, is used to alleviate volumetric locking associated with nearly incompressible property of elastomers. The proposed framework is validated through full-scale simulations and homogenized response involving composites with varying inclusion volume fractions. Numerical investigations demonstrate that the homogenized models is able to reproduce the global force-displacement response within 2\% error and damage evolution characteristics of the heterogeneous composites while significantly reducing the computational cost.
		\end{abstract}
		
		
		
		
	\end{frontmatter}
 \pagenumbering{arabic}
 \section{Introduction}\label{S1} 
Particulate elastomeric composites, in which reinforcing fillers such as carbon black and silica are dispersed within a rubber matrix, are widely used in engineering applications including tire compounds, seismic isolation bearings, and flexible armors owing to their enhanced mechanical performance~\cite{Lopez-Pamies2013,Guo2014,Michel2010}. Under large deformations, however, the material response is influenced by progressive damage arising from stress concentrations within the matrix and at matrix--reinforcement interfaces due to the contrast in constituent properties. The resulting stiffness degradation and strain localization, together with the nonlinear response of elastomers and the multiscale nature of damage evolution, make the prediction of damage in particulate elastomeric composites challenging~\cite{Bleiler2019,Clment2012,Ryu2019,Dorhmi2020,Ma2024,Pea2014,Crespo2020}.

The explicit resolution of the heterogeneous microstructure in structural-scale simulations can result in substantial computational cost, particularly when nonlinear deformation and progressive damage are considered. An effective constitutive description that accounts for the influence of the constituent phases while avoiding explicit resolution of the complete microstructure is therefore desirable for structural-scale analysis. The prediction of the effective constitutive response of nonlinear heterogeneous materials has been extensively investigated using variational homogenization methods. Based on the variational principles of Hashin and Shtrikman, Willis~\cite{OnteCastaneda1989,DeSouzaNeto2010} introduced a homogeneous comparison medium, in which the heterogeneous composite is approximated by an equivalent homogeneous material characterized by uniform constitutive properties. Ponte Castañeda subsequently generalized this concept by introducing a linear comparison composite (LCC), wherein the heterogeneous microstructure is retained while the nonlinear constitutive response of each constituent is replaced with it's linearized representation. This formulation led to the second-order homogenization method, which provides more closer estimates of the effective response of nonlinear composites while maintaining computational efficiency~\cite{PonteCastaeda2000,PonteCastaeda2011,PonteCastaeda2012}. The framework has been successfully applied to hyperelastic composites~\cite{Bleiler2021,Tian2021,Bleiler2019,Lahellec2004,Bouchart2008,Bouchart2010}, elasto-viscoplastic composites~\cite{Doghri2010,Geers2017}, elasto-plastic composites~\cite{Brassart2011}, composite polymer melts~\cite{Domurath2012,Guo2014}, carbon-black-filled elastomers~\cite{PonteCastaeda2000,Zhou2013}, nanocomposites~\cite{Allegra2008,Kari2007}, and particle-reinforced elastomers~\cite{Avazmohammadi2013,Yao2022}. In particular, Bouchart \textit{et al.}~\cite{Bouchart2008,Bouchart2010} implemented and numerically investigated the second-order homogenization framework for hyperelastic composites with particulate inhomogeneities, providing benchmark results for assessing the predictions of the homogenization approach.

The nearly incompressible behavior of elastomeric composites has also been addressed within homogenization frameworks. Ponte Castañeda~\cite{PonteCastaeda2000} considered the limiting case of complete incompressibility for particulate elastomeric composites with rigid inhomogeneities. Avazmohammadi \textit{et al.} ~\cite{Avazmohammadi2013,Clment2012,Clment2013} considered nearly incompressible elastomeric composites with particulate and cylindrical inhomogeneities using a volumetric--deviatoric decomposition of the
strain-energy density together with an effective bulk modulus.

Various damage modeling strategies have been proposed in the context of micromechanical analysis of heterogeneous materials. Among them, the Gurson model and its extensions have been extensively employed for porous materials, where damage evolution is represented through the growth of void volume fraction and its influence on the effective constitutive response \cite{Dormieux2010}. The same framework has also been utilized to investigate matrix--particle interfacial stress states using idealized cell models \cite{Tekoglu2010}. For particulate composites, additional damage mechanisms such as matrix degradation, particle cracking, and matrix--particle decohesion have motivated the development of coupled damage and homogenization frameworks, including approaches based on the Gologanu--Leblond--Devaux constitutive model and multiscale finite element analyses. Alternative methodologies have been applied for metal matrix composites undergoing manufacturing processes such as rolling and extrusion, where the effective constitutive response is modified using damage measures obtained from  finite element simulations \cite{Dorhmi2020}. More recently, stochastic micromechanical analyses \cite{Wu2016}, phase-field formulations \cite{Ma2024,Pranavi2024}, and full-field multiscale simulations~\cite{Bargmann2018} have been employed to investigate damage initiation and propagation in heterogeneous composites. Although these approaches provide detailed insight into the underlying damage mechanisms, they often require explicit resolution of the microstructure and consequently entail significant computational expense for engineering applications.

Motivated by the need for computationally efficient structural-scale damage simulations, the present work integrates the damage analysis of particulate elastomeric composites using overall properties obtained by a homogenization approach. Recent continuum damage formulations for hyperelastic materials~\cite{Pabbu2025,Comellas2016,Anssari-Benam2022} provide an efficient description of stiffness degradation using a limited number of material parameters. The effective constitutive response obtained from the second-order homogenization method is employed to identify an equivalent homogeneous constitutive model. The effective parameters is then integrated with an isotropic continuum damage formulation for structural-scale simulations. This approach enables damage analysis without explicitly resolving the underlying heterogeneous microstructure. A mixed displacement-pressure formulation is adopted within the finite element framework to account the nearly incompressible behavior of elastomeric materials~\cite{Kadapa2022,Azevedo2006}.

In this paper, Section~\ref{s2_1} presents the constitutive framework for homogeneous materials, Section~\ref{S2} presents the second-order homogenization framework and its validation. Section~\ref{S3} describes the continuum damage formulation and its finite element implementation within a mixed displacement--pressure framework. Subsequently, numerical examples involving heterogeneous and homogenized particulate elastomeric composites with varying inclusion volume fractions are presented to assess the accuracy and computational efficiency of the proposed methodology. Finally, the main findings and conclusions of the study are summarized in Section~\ref{S4}.

\section*{Notation}
Throughout this work, a standard tensor notation is employed. Scalar quantities
are denoted by light-face Latin and Greek letters (e.g., $J$, $\psi$, and
$\tau$), vectors by bold lower-case Latin letters (e.g., $\boldsymbol{x}$),
second-order tensors by bold upper-case Latin letters (e.g.,
$\boldsymbol{F}$, $\boldsymbol{P}$, $\boldsymbol{S}$, and $\boldsymbol{I}$),
and fourth-order tensors by blackboard bold symbols (e.g.,
$\mathbb{C}$, $\mathbb{P}$, $\mathbb{S}$, and $\mathbb{A}$). Tensor
components are denoted by lower-case Latin indices, with repeated indices
implying summation. The transpose of a second-order tensor is denoted by
$\left( \bullet \right)^{\mathrm{T}}$, such that
${A}^{\mathrm{T}}_{ij}=A_{ji}$. The inner product or projection of a second-order tensor is indicated as $\left(\blacksquare\right)\cdot(\bullet)$, which in index notation is given as $(\blacksquare)_{ij}(\bullet)_{ij}$. The outer product between two
second-order tensors is denoted by ``$\otimes$'', with its components defined
as $\left(\blacksquare\otimes\bullet\right)_{ijkl} = \left(\blacksquare\right)_{ij}\left(\bullet\right)_{kl}$. The
Kronecker product between two second-order tensors is denoted by ``$\times$'',
with its components defined as
$\left(\blacksquare\times\bullet\right)_{ijkl}=\left(\blacksquare\right)_{ik}\left(\bullet\right)_{jl}$. A fourth-order tensor, $\mathbb{A}$ linearly transforms a second-order tensor $\boldsymbol{B}$ which is denoted by $\mathbb{A}[\boldsymbol{B}]$, with $\mathbb{A}[\boldsymbol{B}]=A_{ijkl}B_{kl}\boldsymbol{e}_{i}\otimes\boldsymbol{e}_{j}$.
The major symmetry and right-minor symmetry of a fourth-order tensor is denoted as $\left(\bullet\right)^\mathrm{T}$ and $\left(\bullet\right)^{\mathrm{T}_{\mathrm{R}}}$ respectively. Major symmetry is defined as $\left(\bullet\right)^\mathrm{T}_{ijkl}=\left(\bullet\right)_{klij}$ and the right-minor symmetry is defined as
$\left(\bullet\right)^{\mathrm{T}_{\mathrm{R}}}_{ijkl}=\left(\bullet\right)_{ijlk}$. The second-order identity tensor is defined as $\boldsymbol{I}=\delta_{ij}\boldsymbol{e}_i\otimes \boldsymbol{e}_j$,
where $\delta_{ij}$ denotes the Kronecker delta. The fourth-order identity tensor, $\mathbb{I}$ is defined
as $\mathbb{I}=\boldsymbol{I}\times\boldsymbol{I}$, while the symmetric
fourth-order identity tensor, $\mathbb{I}^{\mathrm{s}}$ is defined as
$\mathbb{I}^{\mathrm{s}} =\frac{1}{2}(\boldsymbol{I}\times\boldsymbol{I} +(\boldsymbol{I}\times\boldsymbol{I})^\mathrm{T_R})$ . The phase-dependent quantities are identified by the subscript $\gamma$ and the phase volume
fraction is denoted by $c_{\gamma}$. The phase averages are represented by
$\langle\bullet\rangle_{\gamma}$, whereas $\langle\bullet\rangle$ denotes the volume average over the whole heterogeneous composite. The operator $\operatorname{stat}(\bullet)$ denotes the stationary value of the variational functional. 
\\
\section{Constitutive framework for homogeneous hyperelastic materials}\label{s2_1}
A homogeneous hyperelastic body occupying the reference configuration $\Omega_{\mathrm{o}}$ is considered. Under the action of prescribed essential and natural boundary conditions, the body undergoes a motion described by the deformation mapping function $\boldsymbol{\varphi}:\Omega_{\mathrm{o}}\times[0,T]\rightarrow\Omega_{\mathrm{t}}$, where $\Omega_{\mathrm{t}}$ denotes the deformed configuration at time $t$. The  deformation from the material point $\boldsymbol{X}\in\Omega_{\mathrm{o}}$ to its current position $\boldsymbol{x}\in\Omega_{\mathrm{t}}$ is described as
\begin{equation}\label{eq1}
\boldsymbol{x}=\boldsymbol{\varphi}(\boldsymbol{X},t).
\end{equation}
The unsymmetric deformation gradient tensor, which characterizes the  deformation of the material, is defined as
\begin{equation}\label{eq2}
\boldsymbol{F}=\frac{\partial\boldsymbol{x}}{\partial\boldsymbol{X}}
=\boldsymbol{I}+\nabla_{\mathrm{o}}\boldsymbol{u},
\end{equation}
where $\boldsymbol{u}=\boldsymbol{x}-\boldsymbol{X}$ denotes the displacement
vector and $\nabla_{\mathrm{o}}(\bullet)$ represents the gradient operator
with respect to the reference configuration. The symmetric right and left Cauchy--Green
deformation tensors, $\boldsymbol{C}$ and $\boldsymbol{B}$, are defined,
respectively, by
\begin{equation}
\boldsymbol{C}=\boldsymbol{F}^{\mathrm{T}}\boldsymbol{F},
\qquad
\boldsymbol{B}=\boldsymbol{F}\boldsymbol{F}^{\mathrm{T}}.
\end{equation}

For a hyperelastic material, the constitutive behavior is characterized by the strain-energy density function $\psi$, which represents the stored energy per unit reference volume and is expressed as a function of $\boldsymbol{F}$. The constitutive response is obtained by differentiating the strain-energy density function with respect to the deformation measures. The symmetric second Piola--Kirchhoff stress tensor $\boldsymbol{S}$, the first Piola-Kirchhoff (nominal) stress tensor $\boldsymbol{P}$, the fourth-order material tangent tensor $\mathbb{C}$, and the nominal elasticity tensor $\mathbb{A}$ are defined as
\begin{subequations}\label{eq:constitutive_relations}
\begin{align}
\boldsymbol{S}
&=
2\frac{\partial\psi}{\partial\boldsymbol{C}},
\label{eq:S_definition}
\\
\boldsymbol{P}
&=
\frac{\partial\psi}{\partial\boldsymbol{F}},
\label{eq:P_definition}
\\
\mathbb{C}
&=
2\frac{\partial\boldsymbol{S}}{\partial\boldsymbol{C}},
\label{eq:C_definition}
\\
\mathbb{A}
&=
\frac{\partial\boldsymbol{P}}{\partial\boldsymbol{F}}.
\label{eq:A_definition}
\end{align}
\end{subequations}
The term $\mathbb{C}$ and $\mathbb{A}$ are related through $\boldsymbol{F}$ and $\boldsymbol{S}$ which is given as
\begin{equation}
A_{ijkl}
=
\delta_{ik}S_{lj}
+
F_{im}\,C_{mjnl}\,F_{kn},
\end{equation}
The tensor $\mathbb{C}$ possesses both major and minor symmetries, i.e.,
$C_{ijkl}=C_{klij}$, $C_{ijkl}=C_{jikl}$, and $C_{ijkl}=C_{ijlk}$, provided that the $\psi$ function is sufficiently smooth. In contrast, $\mathbb{A}$ generally does not possess the minor symmetry, although its major symmetry follows from its derivation from the $\psi$. The stress--strain pairs $(\boldsymbol{S},\boldsymbol{C})$ and $(\boldsymbol{P},\boldsymbol{F})$ are work-conjugate in the reference configuration, with $\boldsymbol{P}=\boldsymbol{F}\boldsymbol{S}$.
\\
\section{Homogenization of Elastomeric Composites}\label{S2}
 Consider a microstructure of a particulate composite occupying the reference configuration $\Omega_{\mathrm{o}}$ and comprising $N$ distinct material phases with randomly distributed spherical inhomogeneities, resulting in a statistically isotropic microstructure. The overall behaviour of the composite is characterized by the effective deformation gradient, $\overline{\boldsymbol{F}}$ and the effective strain-energy density function, $\overline{\psi}$. It is assumed that there exists a length scale separation and the equality of macroscopic work and microscopic work (Hill-Mandel condition) must hold. At finite strains, this condition is formulated in terms of  $\boldsymbol{F}$ and $\boldsymbol{P}$, as they satisfy the averaging theorem $\left( \overline{\boldsymbol{F}}=\langle\boldsymbol{F}\rangle,  \overline{\boldsymbol{P}}=\langle\boldsymbol{P}\rangle \right)$. However, nonlinear strain measures generally do not hold the averaging theorem due to their nonlinear dependence on $\boldsymbol{F}$~\cite{Geers2017}.

 The $\overline{\psi}$ of a nonlinear heterogeneous composite is difficult to estimate because of strong fluctuations of local $\boldsymbol{F}$. In view of this difficulty, the second-order tangent homogenization method proposed by Ponte Castañeda~\cite{PonteCastaeda2000} employs a fictitious LCC to obtain an approximation of $\overline{\psi}$. The local strain-energy density of the heterogeneous composite is expressed as
\begin{equation}
\psi(\boldsymbol{X},\boldsymbol{F})=\sum_{\gamma=1}^{N}\mathcal{J}_{\gamma}(\boldsymbol{X})\,\psi_\gamma(\boldsymbol{F})
\end{equation}
where $\mathcal{J}_{\gamma}(\boldsymbol{X})$ denotes the indicator function for $\gamma$ phase, and $\psi_\gamma(\boldsymbol{F})$ represents the phase-specific strain-energy density function. The $\psi_\gamma$ satisfies the objectivity condition $\psi_\gamma(\boldsymbol{Q}\boldsymbol{F})=\psi_\gamma(\boldsymbol{F})$ for every proper orthogonal tensor $\boldsymbol{Q}$. By using polar decomposition, $\overline{\boldsymbol{F}}=\overline{\boldsymbol{R}}\overline{\boldsymbol{U}}$, where $\overline{\boldsymbol{R}}$ and $\overline{\boldsymbol{U}}$ denote the macroscopic rotation and right stretch tensors, respectively. By objectivity, the  $\overline{\boldsymbol{R}}$ can be removed without altering the $\overline{\psi}$, allowing the constitutive response to be evaluated with respect to the  $\overline{\boldsymbol{U}}$. The local deformation fields that actually contribute to  $\overline{\psi}$ are accounted consistently such that their macroscopic average satisfies the prescribed deformation gradient.

The $\overline{\psi}$ is defined as the stationary value of the volume-averaged local strain-energy density over the set of kinematically admissible deformation fields~\cite{PonteCastaeda2000}. In the second-order homogenization framework, the extremum operation of the conventional variational formulation is replaced by a stationary operation (see ~\ref{Appenidx A}), whereby the first variation of the corresponding functional vanishes. This stationary formulation leads to the
second-order estimate of $\overline{\psi}$. The effective potential can be
expressed as
\begin{equation}\label{eq6}
\overline{\psi}(\overline{\boldsymbol{F}})=\operatorname*{stat}_{\boldsymbol{F}\in\mathcal{K}(\overline{\boldsymbol{F}})}\left\langle
\psi(\boldsymbol{X},\boldsymbol{F})
\right\rangle
=
\operatorname*{stat}_{\boldsymbol{F}\in\mathcal{K}(\overline{\boldsymbol{F}})}
\sum_{\gamma=1}^{N}
c_{\gamma}
\left\langle
\psi_\gamma\!\left(\boldsymbol{F}_\gamma\right)
\right\rangle
\end{equation}
where $\mathcal{K}(\overline{\boldsymbol{F}})$ denotes the set of kinematically admissible deformation fields that satisfy the prescribed boundary conditions through $\overline{\boldsymbol{F}}$.

Since the stationary variational problem in Eq.~(\ref{eq6}) generally does not admit a closed-form solution for nonlinear hyperelastic composites, the direct evaluation of the  $\overline{\psi}$ is not feasible. The objective is to estimate the $\overline{\psi}$ of the actual heterogeneous composite, characterized locally by $\psi(\boldsymbol{X},\boldsymbol{F})$. To facilitate this estimation, LCC with strain-energy density $\psi_{\mathrm{o}}(\boldsymbol{X},\boldsymbol{F})$ is introduced as a reference problem. The strain-energy density function assuming an inhomogeneous $\boldsymbol{F}$ in the LCC problem is defined as 
\begin{equation}
\psi_{\mathrm{o}}(\boldsymbol{X},\boldsymbol{F})
=
\frac{1}{2}
\,\boldsymbol{F}\cdot
\mathbb{A}_{\mathrm{o}}(\boldsymbol{X})
\,\left[\boldsymbol{F}\right].
\end{equation}
The local stiffness tensor of the LCC is expressed as
\begin{equation}
\mathbb{A}_{\mathrm{o}}(\boldsymbol{X})
=
\sum_{\gamma=1}^{N}
\mathcal{J}_{\gamma}(\boldsymbol{X})
\,\mathbb{A}_{\mathrm{o}_\gamma}
\end{equation}
where $\mathbb{A}_{\mathrm{o}_\gamma}$ denotes the homogeneous reference stiffness tensor of phase $(\gamma)$. The reference stiffness $\mathbb{A}_{\mathrm{o}_\gamma}$ satisfies the major symmetry condition but is not restricted by the minor symmetry. The difference between the actual and comparison energies is defined as
\begin{equation} \label{eq10}
\Delta\psi(\boldsymbol{X},\boldsymbol{F})
=
\psi(\boldsymbol{X},\boldsymbol{F})
-
\psi_{\mathrm{o}}(\boldsymbol{X},\boldsymbol{F}).
\end{equation}
The polarization tensor $\boldsymbol{T}$ (stress-like term) is defined as the derivative of this energy difference with respect to the deformation gradient,
\begin{equation} 
\boldsymbol{T}
=
\frac{\partial\Delta\psi}{\partial\boldsymbol{F}}
=
\frac{\partial(\psi-\psi_{\mathrm{o}})}
{\partial\boldsymbol{F}},
\end{equation}
and therefore characterizes the constitutive difference between the actual composite and the LCC. Similarly, using the complimentary energy function, the homogeneous $\boldsymbol{F} $ is given as
\begin{equation}
\boldsymbol{F}
=
\frac{\partial\Delta\psi^{*}}
{\partial\boldsymbol{T}},
\end{equation}
where ${\Delta\psi^*}$ is a complementary function which can be obtained by Legendre transformation of Eq. \ref{eq10}
The introduction of the LCC reformulates the stationary variational problem in Eq.~(\ref{eq6}) as
\begin{equation}\label{eq12}
\overline{\psi}(\overline{\boldsymbol{F}})
=
\operatorname*{stat}_{\boldsymbol{T}}
\left\{
\underbrace{
\operatorname*{stat}_{\boldsymbol{F}\in\mathcal{K}(\overline{\boldsymbol{F}})}
\left(
\boldsymbol{T}\cdot\boldsymbol{F}
+
\psi_{\mathrm{o}}(\boldsymbol{X},\boldsymbol{F})
\right)
}_{\displaystyle \psi_{\uptheta}}
-
\left\langle
\Delta\psi^{*}(\boldsymbol{X},\boldsymbol{T})
\right\rangle
\right\}
\end{equation}
The first term in the Eq.(~\ref{eq12}) refers to a fictitious linear thermoelastic energy (see \ref{Appenidx A}, \ref{eqA8}). Following to the derivations in \ref{Appendix B}, the $\overline{\psi}$ can be expressed as
\begin{equation}\label{eq13}
\overline{\psi}(\overline{\boldsymbol{F}})
=\sum_{\gamma=1}^{N}c_{\gamma}
\left\{
\psi_\gamma\!\left(\overline{\boldsymbol{F}}_\gamma\right)
+
\frac{1}{2}
\left(
\boldsymbol{F}
-
\overline{\boldsymbol{F}}_\gamma
\right)
\cdot
\frac{\partial \psi_\gamma
\!\left(\overline{\boldsymbol{F}}_\gamma\right)}
{\partial \boldsymbol{F}}\right\}.
\end{equation}
The effective nominal stress, $\overline{\boldsymbol{P}}$ is obtained by differentiating the $\overline{\psi}$ with respect to the $\overline{\boldsymbol{F}}$ and it is expressed as
\begin{equation}\label{eq17}
    \begin{aligned}
        \overline{\boldsymbol{P}}
        &=
        \left\langle \boldsymbol{P} \right\rangle
        =
        \frac{\partial \overline{\psi}\!\left(\overline{\boldsymbol{F}}\right)}
        {\partial \overline{\boldsymbol{F}}}\\
        &=\frac{1}{2}\sum_{\gamma=1}^{N}c_\gamma
\Bigg(
\boldsymbol{P}_\gamma
+
\left(\frac{\partial
\overline{\boldsymbol{F}}_\gamma}
{\partial\overline{\boldsymbol{F}}}\right)^\mathrm{T} \Big[
\boldsymbol{P}_\gamma
+
\mathbb{A}_{\gamma}
\left[
\overline{\boldsymbol{F}}
-
\overline{\boldsymbol{F}}_\gamma
\right]
\Big]
\Bigg).
    \end{aligned}
\end{equation}

The term $\displaystyle \frac{\partial \overline{\boldsymbol{F}}_{\gamma}}{\partial \overline{\boldsymbol{F}}}$ implicitly accounts for the local fluctuation fields through concentration relations, providing a means to evaluate the effective response (see Eq.~(\ref{eq6})). The implementation of the second-order estimate requires the  $\overline{\boldsymbol{F}}_\gamma$ and  $\displaystyle \frac{\partial \overline{\boldsymbol{F}}_{\gamma}}{\partial \overline{\boldsymbol{F}}}$, which are to be computed by solving a linear thermoelastic homogenization problem.
In a small-strain framework, the phase-average strain in a two-phase linear thermo-elastic problem can be related
to the applied macroscopic strain and the difference in the fictitious thermal stress which is given as ~\cite{pallicity2025statisticsresidualstressrandom,LIELENS199863}
\begin{equation}\label{eq18}
\overline{\boldsymbol{\varepsilon}}_\gamma
=
\mathbb{B}_\gamma
\left[\boldsymbol{E}\right]
+
\left(
\mathbb{B}_\gamma-\mathbb{I}^\mathrm{s}
\right)
\left(
\Delta\mathbb{C}
\right)^{-1}
\left[
\Delta\boldsymbol{\beta}
\right],
\end{equation}
where $\mathbb{B}_\gamma$ is the strain concentration tensor of the phase
$\gamma$, $\mathbb{I}^\mathrm{s}$ denotes the symmetric fourth-order identity tensor,
$\Delta\mathbb{C}$ is the stiffness difference $\left(\mathbb{C}_2-\mathbb{C}_1, \mathbb{C}_2>\mathbb{C}_1\right)$, and
$\Delta\boldsymbol{\beta}$
is the thermal stress difference. 
For the finite-deformation hyperelastic problem considered here, the constitutive response is nonlinear and is therefore treated locally through an incremental (linearized) formulation. Following the linearized treatment of hyperelastic composites~\cite{Das2021}, in an analogous manner the phase-average deformation gradient $\overline{\boldsymbol{F}}_\gamma$ can be expressed as 
\begin{equation}\label{eq18}
\overline{\boldsymbol{F}}_\gamma
=
\mathbb{B}_\gamma
\left[\overline{\boldsymbol{F}}\right]
+
\left(
\mathbb{B}_\gamma-\mathbb{I}
\right)
\left(
\Delta\mathbb{A}
\right)^{-1}
\left[
\Delta\boldsymbol{T}
\right],
\end{equation}
where $\mathbb{B}_\gamma$ is the concentration tensor of phase
$\gamma$, $\mathbb{I}$ denotes the fourth-order identity tensor,
$\Delta\mathbb{A}$ is the stiffness difference, and
$\boldsymbol{T}_\gamma
=
\boldsymbol{P}_\gamma
-
\mathbb{A}_{\mathrm{o}_\gamma}
\left[\overline{\boldsymbol{F}}_\gamma\right]$
is the polarization stress.
For a two-phase particulate composite with phase volume fractions $c_1$ and $c_2$, the effective stiffness using Walpole notation is given as,
\begin{equation}\label{}
    \begin{aligned}
\overline{\mathbb{A}}
=
\langle \mathbb{L} \rangle^{-1} + {\mathbb{A}}_{*}-\mathbb{P}^{-1}  \\
\mathbb{L}=\left(\mathbb{A}-\mathbb{A}_{*} +\mathbb{P}^{-1}\right)^{-1}
    \end{aligned}
\end{equation}
In a 2-phase composite, the concentration tensor of the matrix phase is defined from the effective stiffness using the relation,
\begin{equation}\label{eq19}
    \begin{aligned}
    \mathbb{B}_1=\frac{1}{c_1}\left(\Delta\mathbb{A}\right)^{-1}\left(\mathbb{A}_2-\overline{\mathbb{A}}\right) \\
    \langle\mathbb{B}\rangle= c_1\mathbb{B}_1+ c_2\mathbb{B}_2=\mathbb{I}
    \end{aligned}
\end{equation}
The Hill's polarization tensor $\mathbb{P}$ depends on the inhomogeneity shape and the choice of reference stiffness in a linear homogenization problem. It is given as
\begin{equation}\label{eq20}
    \begin{aligned}
        \mathbb{P}=\frac{1}{4\pi}\int_{0}^{2\pi}\int_{0}^{\pi}\mathbb{H}\,\sin\theta \, \mathrm{d}\theta \,\mathrm{d}\varphi\\
        H_{ijkl}=K_{ik}^{-1}\,n_j\,n_l\\
        K_{ik}=A^{(*)}_{ijkl}\,n_j\,n_l
    \end{aligned}
\end{equation}
Here, $\left[\boldsymbol{n}\right]
=
\left[
\sin\theta\cos\varphi,\,
\sin\theta\sin\varphi,\,
\cos\theta
\right]^{\mathrm{T}}$ is the unit normal vector, $\boldsymbol{K}$ is the Christoffel tensor with symmetry, and  $\mathbb{A}_{*}$ is the choice of the reference stiffness, which is taken as the stiffness of the matrix phase.  
\subsection{Implementation}
\begin{figure}[htbp]
    \centering
    \includegraphics[height=0.75\textheight]{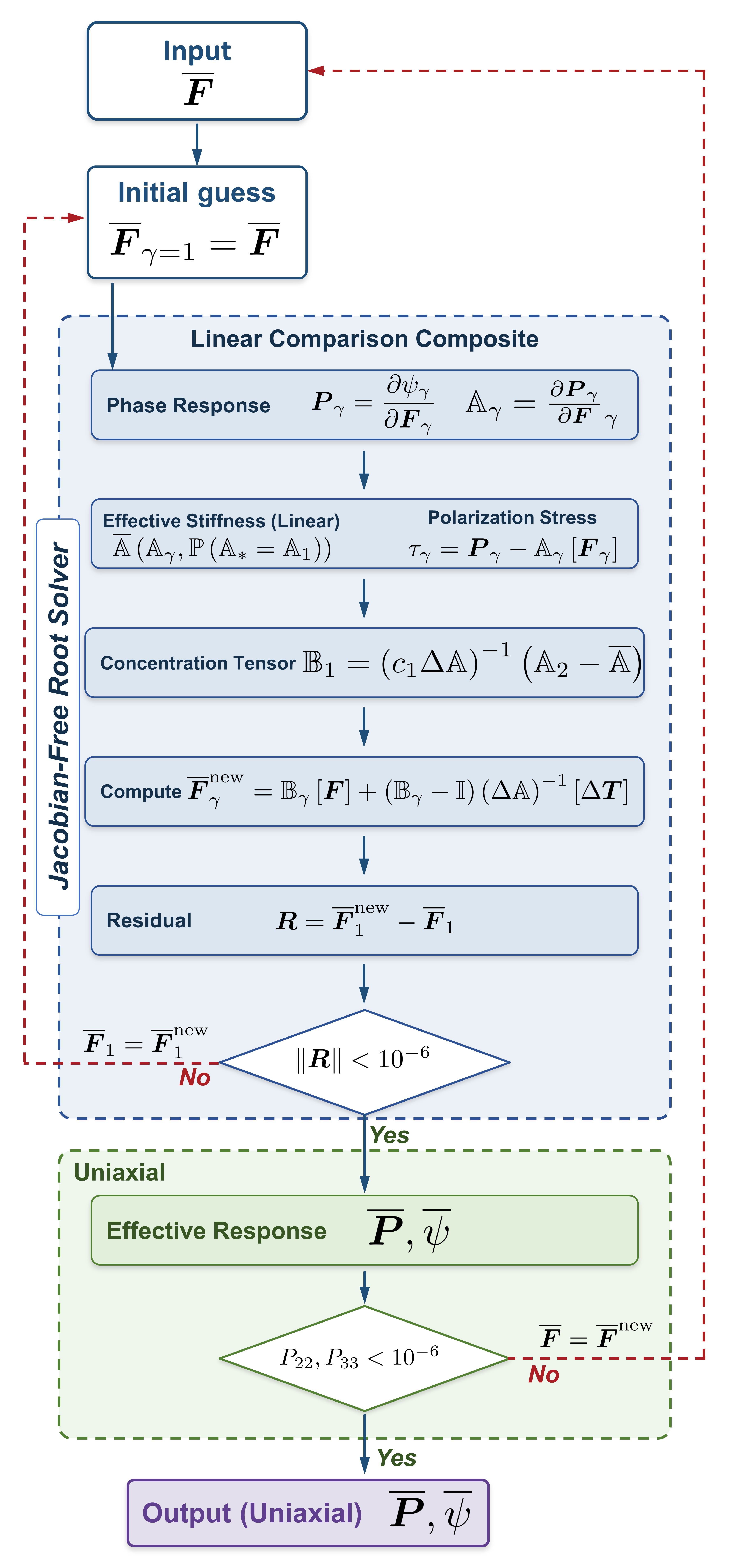}
    \caption{
    Schematic of the workflow depicting the second-order homogenization procedure. For each loading increment, $\overline{\boldsymbol{F}}$ is input. The phase constitutive response, polarization stresses, Hill tensor, and strain localization tensors are evaluated iteratively through the LCC. Upon convergence, the $\overline{\psi}$ and homogenized stress tensor $\boldsymbol{\overline{P}}$ are obtained. Uniaxial state of stress is obtained by ensuring $\overline{P}_{22}=\overline{P}_{33}=0$. The resulting $\overline{\psi}$ and $\boldsymbol{\overline{P}}$ represents for uniaxial tensile test.
    }
    \label{fig:algorithm}
\end{figure}

The implementation of the second-order homogenization framework is schematically shown in Figure ~\ref{fig:algorithm}. The algorithm computes the phase constitutive response, constructs the LCC, evaluates the strain concentration tensors, and subsequently determines the effective strain-energy density and homogenized constitutive quantities using Eqs.~(\ref{eq17})--(\ref{eq20}). The resulting stationary problem is solved using a gradient-free solver, as an analytical expression of the gradient is not possible due to interlinked relations of Eq.(\ref{eq18})-(\ref{eq20}) to find Eq.(\ref{eq17}). Evaluation of the effective quantities requires the evaluation of $\mathbb{P}$ and $\partial \overline{\boldsymbol{F}}_{\gamma}/\partial \overline{\boldsymbol{F}}$. Equation (\ref{eq20}) carries out the integration over the unit sphere, which is evaluated numerically. A Gaussian quadrature with 30 integration points in the $\varphi$
direction and 15 integration points in the $\theta$ direction is employed
in the present work, resulting in 450 angular quadrature points~\cite{GIRAUD20071006,GIRAUD20072627}. This
resolution is selected after verifying the convergence of the Hill
polarization tensor with respect to further refinement of the integration points. It is to be noted that the choice of the number of integration points is very sensitive to the shape of the inhomogeneity (for e.g. aspect ratio of an ellipsoid). The term  $\partial \overline{\boldsymbol{F}}_{\gamma}/\partial \overline{\boldsymbol{F}}$, is computed using a central finite-difference scheme. The whole homogenization procedure is implemented and solved in MATLAB. Later, it is validated against available results reported in the literature. 

For elastomeric composites subjected to uniaxial loading, the lateral traction-free conditions must be satisfied throughout the deformation process. At each prescribed loading increment, the macroscopic deformation gradient is iteratively updated until the transverse components of the nominal stress tensor satisfy ${P}_{22}={P}_{33}=0$, within a prescribed numerical tolerance. Upon convergence, the effective constitutive responses are evaluated.\\
\subsection{Validation}\label{validation}
Second-order mean-field and full-field homogenization are carried out for a two-phase particulate composite with spherical inhomogeneities of volume fraction of 0.15 and 0.25. The validation study is based on the results reported by Bouchart \textit{et al.}~\cite{Bouchart2008, Bouchart2010}, which involve uniaxial loading of the particulate composite. The elastomeric matrix is described by the Mooney-Rivlin (MR) constitutive model, whereas the inhomogeneities are assumed to follow a Neo-Hookean (NH) hyperelastic response. The strain-energy density function of the matrix is given by
\begin{equation}
		\psi=C_{10}(\bar{I_{1}}-3)+C_{01}(\bar{I_{2}}-3)+\frac{1}{D_{1}}(J-1)^2
	\end{equation}
where $C_{10}$ and $C_{01}$ are the material parameters, and $\bar{I}_{1}$ and $\bar{I}_{2}$ denote the modified invariants of $\boldsymbol{C}$. Neo-Hookean model can be obtained as a special case by setting $C_{01}=0$. For the purpose of validation, the effective constitutive response is characterized by the $\overline{\boldsymbol{P}}$ and the $\mathbb{A}$. These quantities are obtained from the $\overline{\psi}$ and are subsequently employed for comparison with  results reported in the literature. The expressions for $\boldsymbol{P}$ and $\mathbb{A}$ of Mooney-Rivlin  model are given by
\begin{equation}
    \begin{aligned}
        \boldsymbol{P}&=C_{10}\left[2\boldsymbol{F}J^{\frac{-2}{3}}-\frac{2}{3}I_{1}J^{\frac{-2}{3}}\boldsymbol{F}^{-\text{T}}\right]+C_{01}\left[2I_{1}\boldsymbol{F}J^{\frac{-4}{3}}-2\boldsymbol{F}\boldsymbol{C}J^{\frac{-4}{3}}-\frac{4}{3}I_{2}J^{\frac{-4}{3}}\boldsymbol{F}^{-\text{T}}\right] \notag \\
	&+\frac{2}{D_{1}}(J-1)J\boldsymbol{F}^{-\text{T}}
    \end{aligned}
\end{equation}
and the nominal elasticity tensor is expressed as
\begin{equation}
\begin{aligned}
\mathbb{A}
={}&
\left[
2C_{10}J^{-\frac{2}{3}}
+2C_{01}I_1J^{-\frac{4}{3}}
\right]\mathbb{I}
\\
&+
\left[
-\frac{4}{3}C_{10}J^{-\frac{2}{3}}
-\frac{8}{3}C_{01}I_1J^{-\frac{4}{3}}
\right]
\left(\boldsymbol{F}\otimes\boldsymbol{F}^{-\mathrm{T}}\right)
\\
&+
\left[
-\frac{4}{3}C_{10}J^{-\frac{2}{3}}
-\frac{8}{3}C_{01}I_1J^{-\frac{4}{3}}
\right]
\left(\boldsymbol{F}^{-\mathrm{T}}\otimes\boldsymbol{F}\right)
\\
&+
\left[
\frac{4}{9}C_{10}I_1J^{-\frac{2}{3}}
+\frac{16}{9}C_{01}I_2J^{-\frac{4}{3}}
+\frac{2}{D_1}(2J-1)J
\right]
\left(\boldsymbol{F}^{-\mathrm{T}}\otimes
\boldsymbol{F}^{-\mathrm{T}}\right)
\\
&+
\left[
\frac{2}{3}C_{10}I_1J^{-\frac{2}{3}}
+\frac{4}{3}C_{01}I_2J^{-\frac{4}{3}}
-\frac{2}{D_1}(J^2-J)
\right]
\left(\boldsymbol{F}^{-\mathrm{T}}\times
\boldsymbol{F}^{-\mathrm{T}}\right)^{\mathrm{T}_{R}}
\\
&+
C_{01}J^{-\frac{4}{3}}
\bigg[
4\left(\boldsymbol{F}\otimes\boldsymbol{F}\right)
-2\left(\boldsymbol{I}\times\boldsymbol{C}\right)
-2\left(\boldsymbol{F}\times\boldsymbol{F}\right)^{\mathrm{T}_{R}}
\\[-2pt]
&\hspace{42mm}
-2\left(\boldsymbol{F}\boldsymbol{F}^{\mathrm{T}}
\times\boldsymbol{I}\right)^{\mathrm{T}_{R}}
+\frac{8}{3}
\left((\boldsymbol{F}\boldsymbol{C})
\otimes\boldsymbol{F}^{-\mathrm{T}}\right)
\\[-2pt]
&\hspace{42mm}
+\frac{8}{3}
\left(\boldsymbol{F}^{-\mathrm{T}}\otimes
(\boldsymbol{F}\boldsymbol{C})\right)
\bigg].
\end{aligned}
\label{eq:C_rearranged}
\end{equation}
The detailed derivation of the above tangent stiffness is available in ~\ref{Appendix B}.
 The parameters for the Mooney-Rivlin model are $C_{10}=C_{01}=1~\text{MPa}$ and $D_{1}=2/3~\text{MPa}^{-1}$, while the parameters for a Neo-Hookean model are $C_{10}=5~\text{MPa}$ and $D_{1}=0.8~\text{MPa}^{-1}$. \\

\begin{figure}[H]
     \centering
     \includegraphics[width=0.7\textwidth]{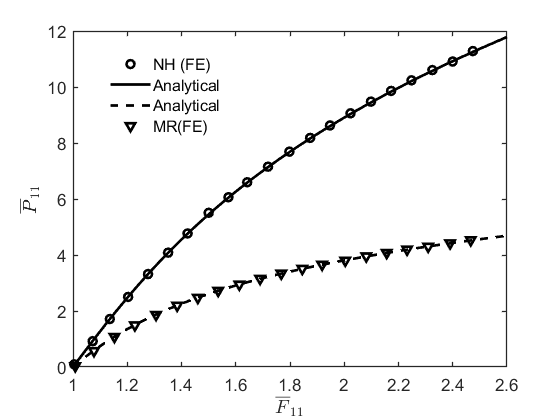}
     \caption{Verification of the implementation of tangent stiffness in FORTRAN for uniaxial loading of the homogeneous material. Comparison of the analytical and FE responses in terms of $\overline{P}_{11}$, as a function of $\overline{F}_{11}$, for the MR and NH models.}
     \label{fig:homogeneous_verification}
 \end{figure}

\begin{figure}[htbp]
    \centering
    \includegraphics[width=0.7\textwidth]{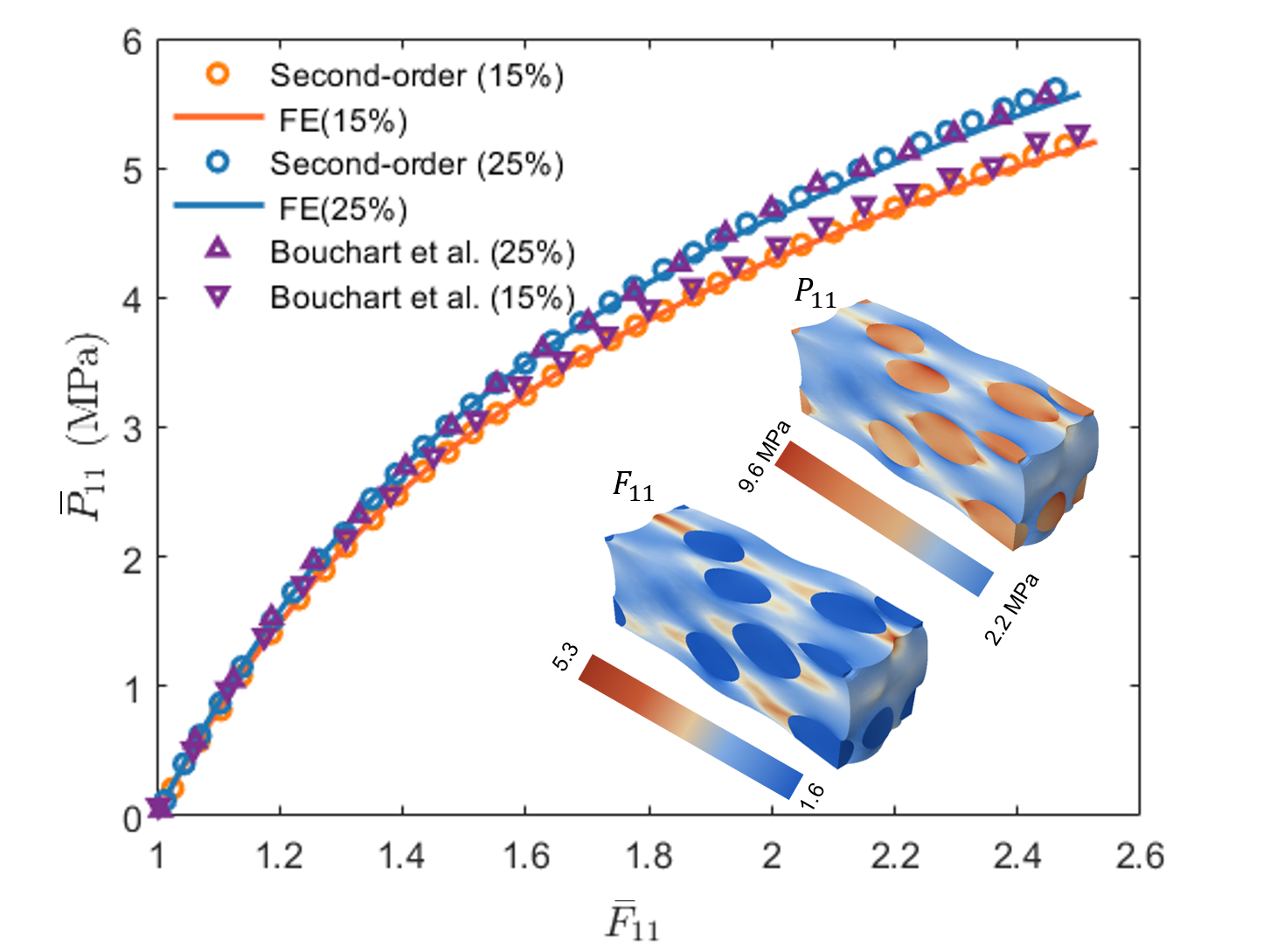}
    \caption{Comparison of the predicted $\overline{\boldsymbol{P}}_{11}-\overline{\boldsymbol{F}}_{11}$ responses with the benchmark results of Bouchart \textit{et al.} for particulate elastomeric composites containing spherical inhomogeneities with volume fractions of $15\%$ and $25\%$.}
    \label{fig:heterogeneous_verification}
\end{figure}

\begin{figure}[H]
     \centering
     \includegraphics[width=0.7\textwidth]{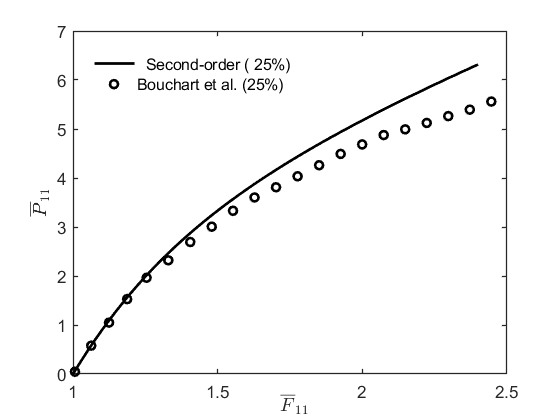}
     \caption{Comparison of $\overline{\boldsymbol{P}}_{11}-\overline{\boldsymbol{F}}_{11}$ response obtained using the numerical evaluation of the Hill polarization tensor and the analytical formulation proposed by Zeka \textit{et al.} for a particulate elastomeric composite subjected to uniaxial loading.}
     \label{fig:eshelby}
 \end{figure}

\begin{figure}[htbp]
    \centering
    \includegraphics[width=0.7\textwidth]{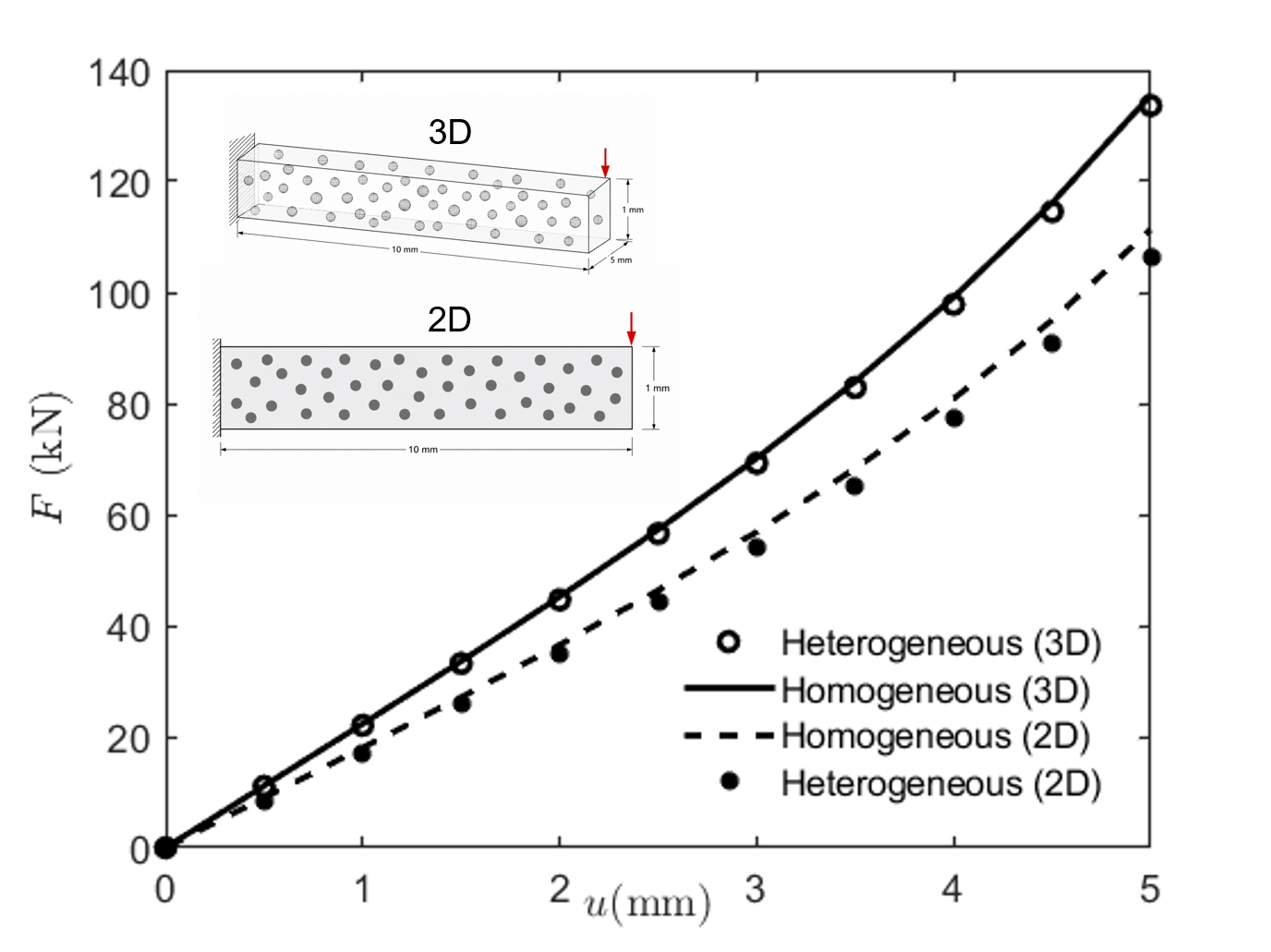}
    \caption{Structural-scale validation of the proposed homogenization
    framework through two- and three-dimensional cantilever beam analyses,
    showing the boundary conditions and comparison of the
    force--displacement responses obtained from the heterogeneous composite
    and the equivalent homogenized models.}
    \label{fig:cantilever_validation}
\end{figure}

A user-defined code is implemented in FORTRAN, compatible with ABAQUS, as it does not yield the field variables ${\boldsymbol{F}}$ and ${\boldsymbol{P}}$ in its output database. As a preliminary verification step, identical constitutive properties are assigned to both the phases, and it is expected that the homogenized response must reproduce the same constitutive behaviour of the underlying material model. Figure~\ref{fig:homogeneous_verification} compares the $\overline{P}_{11}-\overline{F}_{11}$ response obtained from the analytical solution for homogeneous phases characterized by the NH and MR model. The agreement observed in both cases validates the implementation of the constitutive model, and the in-house post-processing codes are independently validated on linear problems.
Following the verification of the homogeneous material response, the second-order homogenization framework is employed to predict the effective behavior of particulate composites. Figure~\ref{fig:heterogeneous_verification} compares the predicted effective nominal stress-stretch response with the benchmark results reported by Bouchart \textit{et al.} under uniaxial loading conditions.
It is observed that they compare well with the reported effective behaviour. Furthermore, a stiffer response is observed for $c_2=0.25$ than for $c_2=0.15$. The FE results indicate significant local variations in \(P_{11}\) within the inhomogeneities, while the \(F_{11}\) field remains nearly homogeneous.  The computation of $\mathbb{P}$ requires evaluating 81 tensor components numerically, making it computationally demanding. In view of this, the isotropic approximation of $\mathbb{P}$ (~\cite{Zeka2022}) significantly reduce the computational cost. However, they exhibit convergence difficulties and increased numerical sensitivity for nearly incompressible case. The effective nominal stress--stretch response is shown in Fig.~\ref{fig:eshelby}.
The effective constitutive parameters of the particulate composites are then obtained by curve-fitting of $\overline{{P}}_{11}-\overline{{F}}_{11}$ data using ABAQUS. The structural performance (force-displacement response) of two- and three-dimensional cantilever beams made of particulate composite is obtained by explicitly modelling the random distribution of inhomogeneities. Later, the effective constitutive parameters are used to model the equivalent homogeneous beams. Figure ~\ref{fig:cantilever_validation} shows that the homogenized models are able to reproduce the structural response both in two- and three-dimensional models at a considerably reduced number of degrees of freedom and computational cost, highlighting their suitability for large-scale simulations involving nonlinear constitutive behaviour and damage evolution.

\section{Continuum Damage Modeling Based on Homogenized Response}\label{S3}
Section 2 established the effective elastic response of undamaged particulate elastomeric composites. Under large deformation, however, these composites undergo progressive stiffness degradation and eventual failure, driven primarily by damage nucleation and growth within the matrix phase, promoted by the stress and strain concentrations that develop in the vicinity of the inhomogeneities. Capturing this behavior requires extending the homogenization framework with a continuum damage formulation, developed in this section and implemented consistently for both the heterogeneous composite and its homogenized counterpart within a mixed displacement–pressure ($\boldsymbol{u}$-$p$) finite element framework. For the damage analysis, the numerical investigations are restricted to two-dimensional configurations to facilitate a systematic assessment of the damage formulation while retaining the essential features of the proposed homogenization framework.\\

Consistent with experimental observations in filled elastomers, damage is assumed to initiate and evolve exclusively within the matrix phase, including matrix elements adjacent to the inclusion boundary, where local stress and strain concentrations are most severe. The stiff inhomogeneities are treated as intrinsically undamaged; particle fracture and fragmentation lie outside the scope of the present formulation. It is emphasized that damage initiation at the matrix–particle interface itself, for example through a cohesive-zone or interfacial decohesion law, is not considered here. All degradation is represented as a bulk continuum phenomenon within the matrix material, even where that material borders an inclusion. Interfacial debonding and cavitation, which are documented damage mechanisms in filled elastomers, therefore remain outside the scope of the present framework and represent a natural direction for future extension.\\

Consider a deformable body occupying the reference configuration $\Omega_{0}$ with boundary $\partial\Omega_{0}=\Gamma_{u}\cup\Gamma_{t}$, where $\Gamma_{u}$ and $\Gamma_{t}$ denote the portions of the boundary subjected to prescribed displacements and tractions, respectively. Let $\bar{\boldsymbol{u}}$ and $\bar{\boldsymbol{t}}$ represent the prescribed displacement and traction fields on $\Gamma_{u}$ and $\Gamma_{t}$. Under quasi-static loading conditions, the equilibrium configuration is characterized by the stationarity of the total potential energy functional, which is composed of the internal strain energy and the potential of the external loads, and is given by
\begin{equation}\label{eq21}
	\begin{aligned}
		\Pi(\boldsymbol{u}) &=\Pi_\text{int}(\boldsymbol{u})-\Pi_\text{ext}(\boldsymbol{u}),\\
		\Pi_\text{int}(\boldsymbol{u}) &=\int_{\Omega_{\mathrm{o}}} \psi(\boldsymbol{E}) \; \text{d}\Omega,\\
		\Pi_\text{ext} &=\int_{\Gamma_t} \boldsymbol{u}\cdot\boldsymbol{t} \; \text{d}\Gamma_{t} + \int_{\Omega_{0}} \boldsymbol{u}\cdot \boldsymbol{b} \;\text{d}\Omega.
	\end{aligned}
\end{equation}
To address the near-incompressibility constraint associated with elastomeric materials, a mixed displacement--pressure ($\boldsymbol{u}$-$p$) formulation is employed within the finite element framework. The governing variational equations and their finite element discretization are described in the following subsection.
\subsection{Mixed Displacement-Pressure Variational Formulation}
The mixed displacement--pressure ($\boldsymbol{u}$--$p$) formulation employed in the present work follows the standard hybrid finite element framework for nearly incompressible hyperelastic solids. By enforcing the stationarity of the total potential energy with respect to the displacement and pressure fields and subsequently linearizing the governing equations, the element equilibrium equations are obtained in the following matrix form
\begin{equation}
\begin{bmatrix}
\boldsymbol{K}_{uu} & \boldsymbol{K}_{up} \\
\boldsymbol{K}_{pu} & \boldsymbol{K}_{pp}
\end{bmatrix}
\begin{Bmatrix}
\Delta \boldsymbol{u} \\
\Delta \boldsymbol{p}
\end{Bmatrix}
=
-
\begin{Bmatrix}
\boldsymbol{R}_{u} \\
\boldsymbol{R}_{p}
\end{Bmatrix}.
\end{equation}
where $\boldsymbol{K}_{uu}$ is the displacement tangent stiffness matrix, $\boldsymbol{K}_{up}$ and $\boldsymbol{K}_{pu}$ are the displacement--pressure coupling matrices, $\boldsymbol{K}_{pp}$ denotes the pressure tangent matrix, and $\boldsymbol{R}_{u}$ and $R_{p}$ represent the displacement and pressure residual vectors, respectively. For the Q1P0 mixed finite element adopted in this work, these quantities are expressed as
\begin{equation}\label{eq23}
\begin{aligned}
\boldsymbol{K}_{uu}
&=
\int_{\Omega_{\mathrm{o}}}
\boldsymbol{B}_\mathrm{L}^\mathrm{T}
\mathbb{C}
\boldsymbol{B}_\mathrm{L}
\, \mathrm{d}V_0
+
\int_{\Omega_{\mathrm{o}}}
\boldsymbol{B}_\mathrm{NL}^\mathrm{T}
\boldsymbol{S}
\boldsymbol{B}_\mathrm{NL}
\, \mathrm{d}V_0,\\
\boldsymbol{K}_{up}
&=
-
\int_{\Omega_{\mathrm{o}}}
\boldsymbol{B}_\mathrm{NL}^\mathrm{T}
\,
J
\boldsymbol{C}^{-1}
\, \mathrm{d}V_0
=
\boldsymbol{K}_{pu}^\mathrm{T},\\
\boldsymbol{K}_{pp}
&=
-
\int_{\Omega_{\mathrm{o}}}
\frac{1}{\kappa}
\, \mathrm{d}V_0,\\
\boldsymbol{R}_{u}
&=
\int_{\Omega_{\mathrm{o}}}
\boldsymbol{B}_\mathrm{L}^\mathrm{T}
\boldsymbol{S}
\, \mathrm{d}V_0
-
\boldsymbol{F}^{\mathrm{ext}},\\
R_{p}
&=
-
\int_{\Omega_{\mathrm{o}}}
\left(
\frac{\partial \psi_{\mathrm{vol}}}{\partial J}
+\bar{p}
\right)
\frac{1}{\kappa}
\, \mathrm{d}V_0.
\end{aligned}
\end{equation}
The global tangent matrix and residual vector are assembled from the element contributions using the standard finite element assembly procedure, and the resulting nonlinear system is solved iteratively using the Newton--Raphson method. In Eqn.~(\ref{eq23}), $\boldsymbol{B}_\mathrm{L}$ denotes the conventional linear strain--displacement matrix and $\boldsymbol{B}_\mathrm{NL}$ denotes the geometric (nonlinear) strain--displacement matrix arising from finite deformations. Furthermore, $\mathbb{C}$ and $\boldsymbol{S}$ correspond to the material elasticity tensor and the second Piola--Kirchhoff stress tensor, respectively. These constitutive quantities are subsequently modified through the damage formulation to capture stiffness degradation, while for the undamaged state ($D=0$) they recover the original hyperelastic response.
\subsection{Damage Evolution Formulation}\label{S3.2}
The isotropic continuum damage model developed by Pabbu \textit{et al.}~\cite{Pabbu2025} is adopted to describe the progressive degradation of the elastomeric matrix. Consistent with the constitutive framework for nearly incompressible hyperelastic materials, damage is assumed to affect only the deviatoric component of the strain-energy density, while the volumetric contribution remains unchanged to preserve the material incompressibility constraint. This assumption is widely adopted for rubber-like materials, where material degradation is primarily associated with distortional deformation. Consequently, the damaged strain-energy density is expressed as
\begin{equation}
\psi = (1-D)\,\psi_{\mathrm{Dev}} + \psi_{\mathrm{Vol}},
\label{eq:damaged_energy}
\end{equation}
The adopted damage formulation is characterized by only three material parameters: the polynomial order $n$, the damage initiation threshold $\tau_{0}^{\mathrm d}$, and the fracture energy $G_{\mathrm f}$ governing the energy dissipation process. The polynomial order $n$ governs the shape of the damage evolution law and serves as an additional material parameter controlling the softening response~\cite{Pabbu2025}. The damage evolution equation is given by
\begin{equation}
\begin{aligned}
    D&=G(\tau)=\frac{1-\frac{\tau_{0}^\text{d}}{\tau}}{1+H},\\
    g(D)&=\frac{(1-D)^n}{(1-D)^n+q(D)},\\
    q(D)&=\log(1+D)=D-\frac{D^2}{2}+\frac{D^3}{3}-...
    \end{aligned}
\end{equation}
It should be noted that the present formulation neglects damage mechanisms associated with volumetric deformation, such as cavitation and matrix--particle interfacial decohesion, which may become significant under highly triaxial stress states encountered near crack tips or in debonding-dominated failure. Consequently, the model is primarily intended to describe distortion-driven stiffness degradation in nearly incompressible elastomeric composites. The incorporation of volumetric damage mechanisms within the homogenization framework remains an important direction for future work.

where $G(\tau)$ denotes the damage evolution function, $D$ is the scalar damage variable, and $\tau_{0}^{\mathrm{d}}$ is the damage initiation threshold. The quantity $\tau=\sqrt{2\psi_{\mathrm{Dev}}}$ represents the damage-driving energy norm, commonly referred to as the $J_{2}$-norm, and is evaluated from the undamaged deviatoric strain-energy density $\psi_{\mathrm{Dev}}$. The parameter $H$, determined from the material fracture energy, governs the rate of damage evolution. Furthermore, $g(D)$ is the degradation function describing the reduction in material stiffness, whereas $q(D)$ is an auxiliary function introduced to ensure a smooth and stable evolution of damage. 
\subsection{Numerical Implementation}
Having established the governing finite element equations and the damage evolution law, the numerical implementation of the proposed framework is now described. Particular emphasis is placed on the constitutive quantities required for the finite element solution procedure, namely the stress and material elasticity tensors in both the undamaged and damaged configurations. The implementation strategy adopted for the heterogeneous composite and the homogenized model is subsequently summarized through a computational flowchart. Finally, a series of representative numerical examples are presented to assess the predictive capability and computational efficiency of the proposed framework.\\

The numerical examples involve both heterogeneous composites and their homogenized counterparts, with the matrix and particle phases modeled using the Neo--Hookean constitutive law. A reinforced rectangular plate containing a pre-existing crack is considered for inclusion volume fractions of $5\%$, $15\%$, and $25\%$. The constitutive relations required for the finite element implementation are summarized below in terms of the second Piola--Kirchhoff stress tensor $\boldsymbol{S}$ and the material elasticity tensor $\mathbb{C}$ for the undamaged and damaged material states.
\begin{equation}\label{eq25}
	\begin{aligned}
	\boldsymbol{S} &=\frac{\partial \psi}{\partial \boldsymbol{E}}=2\left(g(D)\frac{\partial \psi_\text{Dev}}{\partial \boldsymbol{C}}+\frac{\partial \psi_\text{Vol}}{\partial \boldsymbol{C}}\right), \\
	\boldsymbol{S} &=2\left(g(D)C_{10}(J^\frac{-2}{3}\boldsymbol{I}-\frac{1}{3}\bar{I_{1}}\boldsymbol{C}^{-1})-pJ\boldsymbol{C}^{-1}\right).
\end{aligned}
\end{equation}
\begin{equation}\label{eq26}
	\begin{aligned}
		\mathbb{C} =2\frac{\partial \boldsymbol{S}}{\partial \boldsymbol{C}}&=g(D)\mathbb{C}_\text{Dev}+4\left(\frac{\partial \psi_\text{Dev}}{\partial \boldsymbol{C}}\otimes \frac{\partial g(D)}{\partial \boldsymbol{C}}\right)+\mathbb{C}_\text{Vol},\\
		\frac{\partial g(D)}{\partial \boldsymbol{C}} &=\frac{\partial g(D)}{\partial D}\frac{\partial D}{\partial \tau}\frac{\partial \tau}{\partial \boldsymbol{C}},
	\end{aligned}
\end{equation}
where,
\begin{equation}\label{eq27}
	\begin{aligned}
		\frac{\partial g(D)}{\partial D} &=-\frac{(1-D)^{n}(n(1+D)\log(1+D)+(1-D))}{(1-D^2)\left((1-D)^{n}+\log(1+D)\right)^{2}},\\
		\frac{\partial D}{\partial \tau}&=\frac{\tau_{0}^\text{d}}{\tau^{2}(1+H)},\\
		\frac{\partial \tau}{\partial \boldsymbol{C}}&=\frac{1}{\tau}\frac{\partial \psi_\text{Dev}}{\partial\boldsymbol{C}}.
	\end{aligned}
\end{equation}
When $D=0$, Eq.(\ref{eq26}) reduces to the material elasticity tensor ($\mathbb{C}$), which is given as
\begin{equation}\label{eq28}
	\mathbb{C}=\mathbb{C}_\text{Dev}+\mathbb{C}_\text{Vol},
\end{equation}
where,
\begin{equation}\label{eq32}
	\begin{aligned}
		\mathbb{C}_\text{Dev} &=\frac{\partial}{\partial \boldsymbol{C}}\left(\frac{\partial \psi_\text{Dev}}{\partial \boldsymbol{C}}\right)\\ &=4C_{10}\left(\frac{-1}{3}J^\frac{-2}{3}(\boldsymbol{C}^{-1} \otimes \boldsymbol{I})-\frac{1}{3}J^\frac{-2}{3}(\boldsymbol{I} \otimes \boldsymbol{C}^{-1})+\frac{1}{9}\bar{I_{1}}(\boldsymbol{C}^{-1} \otimes \boldsymbol{C}^{-1})+\frac{1}{3}\bar{I_{1}}(\boldsymbol{C}^{-1} \times \boldsymbol{C}^\text{-T})\mathbb{I}_\text{sym} \right), \\
		\mathbb{C}_\text{Vol} &=\frac{\partial}{\partial \boldsymbol{C}}\left(\frac{\partial \psi_\text{Vol}}{\partial \boldsymbol{C}}\right)\\ &= -p\bar{I_{3}}\left((\boldsymbol{C}^{-1} \otimes \boldsymbol{C}^{-1})-(2(\boldsymbol{C}^{-1} \times \boldsymbol{C}^\text{-T} )\mathbb{I}_\text{sym}\right)+\kappa I_3(\boldsymbol{C}^{-1} \otimes \boldsymbol{C}^{-1}).
	\end{aligned}
\end{equation}
The preceding constitutive relations form the basis of the finite element implementation adopted in the present work. During each load increment, the stress tensor and the material elasticity tensor are evaluated
according to the current damage state and subsequently used in the construction of the element residual vectors and tangent stiffness matrices. The finite element implementation follows the damage algorithm developed in
our previous work~\cite{Pabbu2025}, where the detailed computational procedure for the damage evolution and its integration within the finite element framework is presented. In the present study, the same damage
formulation is employed for both the heterogeneous composite and the homogenized model, with the latter requiring an additional homogenization and parameter-identification step prior to the finite element analysis.

The proposed framework is next applied to a series of representative numerical examples involving reinforced elastomeric composites with varying inclusion volume fractions. Particular attention is given to the influence of homogenization on the predicted damage response, force-displacement behavior, and computational efficiency.
\subsection{Numerical Examples}
To demonstrate the applicability of the proposed methodology, numerical simulations are performed on reinforced elastomeric structures containing randomly distributed inhomogeneities. Both heterogeneous composite models and equivalent homogenized models are considered. The examples focus on a pre-cracked rectangular plate subjected to monotonic loading, with inclusion volume fractions of $5\%$, $15\%$, and $25\%$. The predicted force-displacement response, damage evolution, and computational performance are subsequently compared.
\begin{figure}[H]
	\begin{minipage}{.5\textwidth}
		\centering
		\includegraphics[scale=0.4]{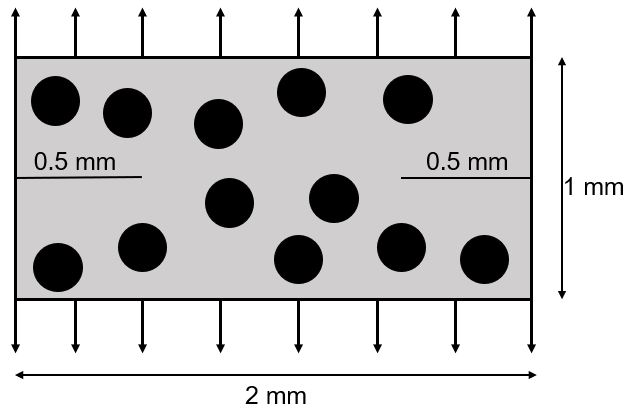}
		\captionof{figure}{Rectangular cracked plate}
  \label{fig9}
	\end{minipage}%
\begin{minipage}{.5\textwidth}
	\centering
	\captionof{table}{Material properties}
	\begin{tabular}{cl}
		\hline
		Property & Value\\
		\hline\rule{0pt}{12pt}
		$C_{10}$  &     7.5 kPa \\
		$\kappa$   &  0.1 GPa \\
		$\tau_{0}^\text{d}$   &  $57.7 ~\text{kPa}$ \\
		$G_\text{f}$   & 20 kN/m \\	[2pt]	
		\hline
	\end{tabular}
\label{T1}
\end{minipage}
\end{figure} 

As shown in Fig.~\ref{fig9}, a two-dimensional rectangular plate of dimensions $(2~\mathrm{mm}\times1~\mathrm{mm})$ containing a double-edge crack of length $(0.5~\mathrm{mm})$ is considered. This benchmark problem is selected to assess the capability of the proposed damage formulation in predicting the initiation and progression of material degradation in the presence of stress concentrations. The plate is subjected to displacement-controlled loading, with one edge fully constrained and a prescribed displacement applied on the opposite edge. The material properties adopted for the matrix phase are summarized in Table~\ref{T1}, while the inclusion properties are chosen such that the stiffness contrast between the particle and matrix phases is maintained at a ratio of 100.

The damage model is applied exclusively to the elastomeric matrix, while the reinforcing inhomogeneities are assumed to behave elastically throughout the analysis. Consequently, a single set of matrix damage parameters is employed uniformly within the matrix phase, and no independent damage evolution is assigned to the inhomogeneities. Particle fracture, fragmentation, and matrix--particle interfacial decohesion are neglected in the present formulation.

\subsubsection{ Composite with 5\% Volume Fraction}
The first example considers a composite containing a $5\%$ inclusion volume fraction. To establish a reference response, the force-displacement behavior of the unreinforced matrix is first examined and compared with that of the heterogeneous composite and its homogenized representation. This comparison enables the influence of the inhomogeneities on the structural response to be quantified while simultaneously assessing the ability of the homogenized model to reproduce the behavior of the heterogeneous composite.
\begin{figure}[H]
     \begin{subfigure}{0.5\textwidth}
     \centering
     \includegraphics[scale=0.5]{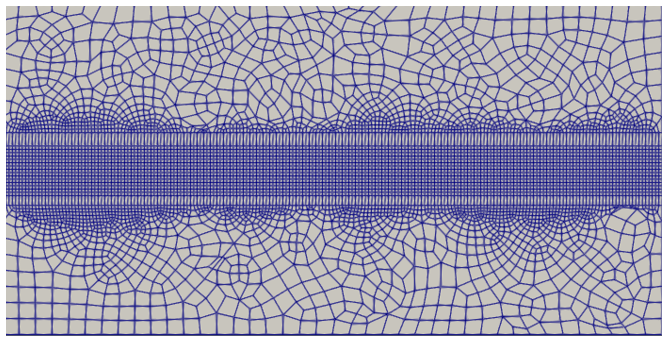}
     \caption{}
     \label{fig10a}
    \end{subfigure}
    \begin{subfigure}{0.5\textwidth}
    	\centering
    	\includegraphics[scale=0.5]{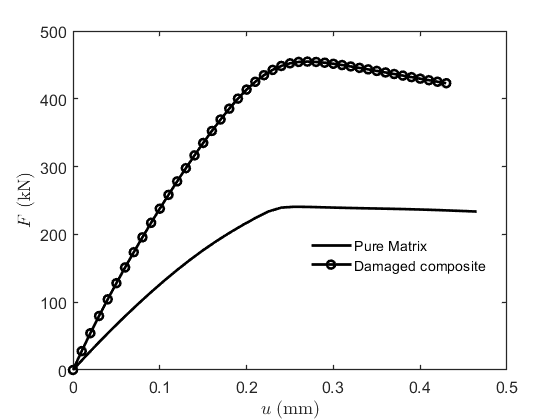}
    	\caption{}
    	\label{fig10b}
    \end{subfigure}
\caption{(a) Finite element mesh adopted for the $5\%$ heterogeneous composite. (b) Comparison of force-displacement responses for the pure matrix, heterogeneous composite.}
\label{fig:pure_matrix}
 \end{figure}
As shown in Fig.~\ref{fig10a}, a converged finite element mesh with an effective element size of $0.025~\mathrm{mm}$ is employed for the heterogeneous composite simulations. Local mesh refinement is introduced in the vicinity of the crack tip to accurately capture the steep stress gradients and damage localization associated with crack propagation while maintaining numerical stability. Consistent with the nearly incompressible behavior of elastomeric materials, plane strain Q1P0 mixed finite elements are adopted throughout the analysis to alleviate volumetric locking. Furthermore, all damage simulations are performed with a polynomial order of $n=1$, since the investigation of different softening responses is beyond the scope of the present study.\\

The force-displacement responses presented in Fig.~\ref{fig10b} indicate that the reinforced composite exhibits a higher load-carrying capacity than the pure matrix material owing to the presence of stiff inhomogeneities. The inclusion phase enhances the overall structural stiffness, resulting in a steeper initial response compared with the unreinforced matrix. Although damage initiation occurs at a relatively earlier stage in the composite due to local stress concentrations generated around the inhomogeneities and crack tip, the reinforcement mechanism enables the structure to sustain higher load levels throughout the deformation process.
\begin{figure}[H]
    \centering
    \includegraphics[width=0.5\linewidth]{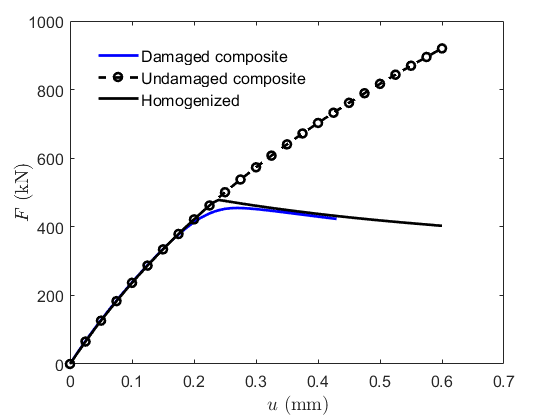}
    \caption{Force--displacement responses of the undamaged composite, damaged heterogeneous composite, and the homogenized model for a $5\%$ inclusion volume fraction}
    \label{fig5percent}
\end{figure}
The force-displacement responses corresponding to the damaged heterogeneous composite, the undamaged composite, and the equivalent homogenized model are presented in Fig.~\ref{fig5percent}. Damage initiation is predicted at a displacement of approximately $u=0.22$-$0.23~\mathrm{mm}$ for both the heterogeneous and homogenized models. The damaged heterogeneous composite attains a peak load of approximately $470~\mathrm{kN}$, whereas the homogenized model predicts a peak load of about $465~\mathrm{kN}$, thereby leading to a difference of nearly $1\%$ only. Furthermore, the post-peak softening response predicted by the homogenized model closely follows that of the heterogeneous composite over the displacement range considered. These results demonstrate that the homogenized constitutive model accurately captures both the onset of damage and the subsequent stiffness degradation of the heterogeneous composite. The close agreement between the two responses demonstrates the capability of the proposed homogenization framework to represent the global mechanical behavior of the heterogeneous structure. The effective hyperelastic material parameters employed in the homogenized model are obtained through a curve-fitting procedure applied to the effective nominal stress-stretch response corresponding to the $5\%$ inclusion volume fraction. The resulting material constants are found to be $C_{10}=8.5~\mathrm{kPa}$ and $\kappa=8.51~\mathrm{MPa}$.\\

It should be noted that the damage initiation threshold $\tau_{0}^{\mathrm{d}}$ employed in the homogenized model cannot be directly adopted from the matrix material, since the presence of inhomogeneities alters the local stress and strain fields, and consequently influences the onset of damage. While the effective hyperelastic constants obtained from the homogenization procedure adequately capture the composite stiffness, they do not directly estimate the damage threshold. To account for this effect, a dimensionless scaling factor $\eta_{\mathrm{d}}$ is introduced to relate the damage initiation threshold of the homogenized model to that of the matrix material:
\begin{equation}
\tau_{0,\mathrm{hom}}^{\mathrm{d}}
=
\eta_{\mathrm{d}}
\tau_{0,\mathrm{m}}^{\mathrm{d}},
\label{eq:hom_damage_threshold}
\end{equation}
where $\tau_{0,\mathrm{m}}^{\mathrm{d}}$ denotes the damage initiation threshold of the matrix material, $\tau_{0,\mathrm{hom}}^{\mathrm{d}}$ denotes the threshold adopted for the homogenized composite with inclusion volume fraction $c$, and $\eta_{\mathrm{d}}$ is the damage-threshold scaling factor accounting for the effect of the inhomogeneities.

The scaling factor is determined from auxiliary simulations performed on uncracked specimens having the same dimensions and material properties as the corresponding cracked configurations. The use of an uncracked geometry
eliminates the influence of the crack-tip stress concentration, thereby enabling the effect of the inhomogeneities on damage initiation to be isolated. For each inclusion volume fraction, the damage-driving quantity $\tau$ is
evaluated within a narrow strip of elements and averaged over all Gauss integration points within the strip. The corresponding average is also evaluated for the pure matrix specimen at the same displacement level associated with the onset of softening. The scaling factor is then defined from the ratio of these two quantities as
\begin{equation}
\eta_{\mathrm{d}}
=
\frac{
\left\langle
\tau
\right\rangle_{\mathrm{strip}}
}{
\left\langle
\tau
\right\rangle_{\mathrm{strip}}^{\mathrm{m}}
},
\label{eq:damage_scaling_factor}
\end{equation}
where $\left\langle\tau\right\rangle_{\mathrm{strip}}$ and $\left\langle\tau\right\rangle_{\mathrm{strip}}^{\mathrm{m}}$ denote the average damage-driving quantity over the selected strip for the composite with volume fraction $c$ and the pure matrix specimen, respectively. For the
volume fractions considered in the present study, the resulting damage initiation thresholds adopted in the homogenized models are $\tau_{0,\mathrm{hom}}^{\mathrm{d}}=0.0452$ MPa, $0.0406$ MPa, and $0.0310$ MPa for inclusion volume fractions of $5\%$, $15\%$, and $25\%$, respectively. The ratio in Eq.~(\ref{eq:damage_scaling_factor}) is
subsequently used in Eq.~(\ref{eq:hom_damage_threshold}) to determine the effective damage initiation threshold employed in the homogenized model.

\begin{figure}[H]
\centering

\begin{subfigure}{0.48\textwidth}
\centering
\includegraphics[width=\linewidth]{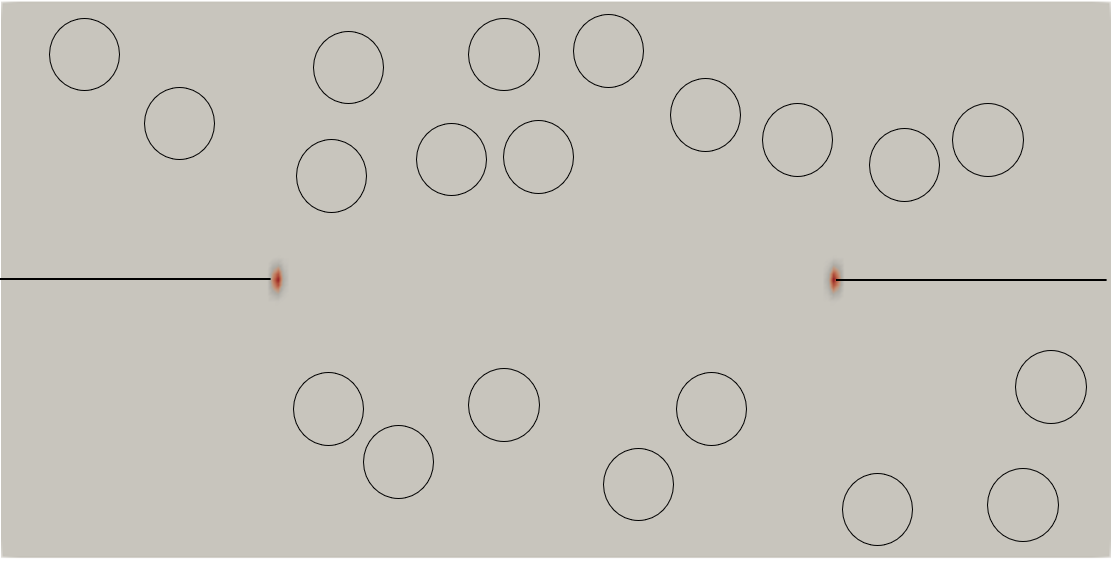}
\caption{}
\end{subfigure}
\hfill
\begin{subfigure}{0.48\textwidth}
\centering
\includegraphics[width=\linewidth]{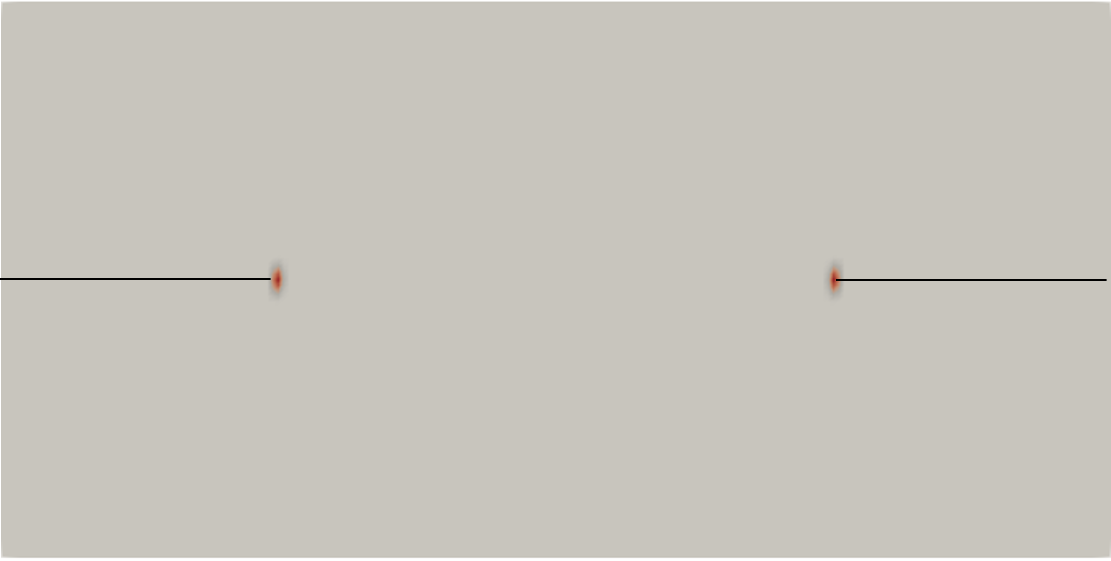}
\caption{}
\end{subfigure}

\vspace{0.2cm}

\begin{subfigure}{0.48\textwidth}
\centering
\includegraphics[width=\linewidth]{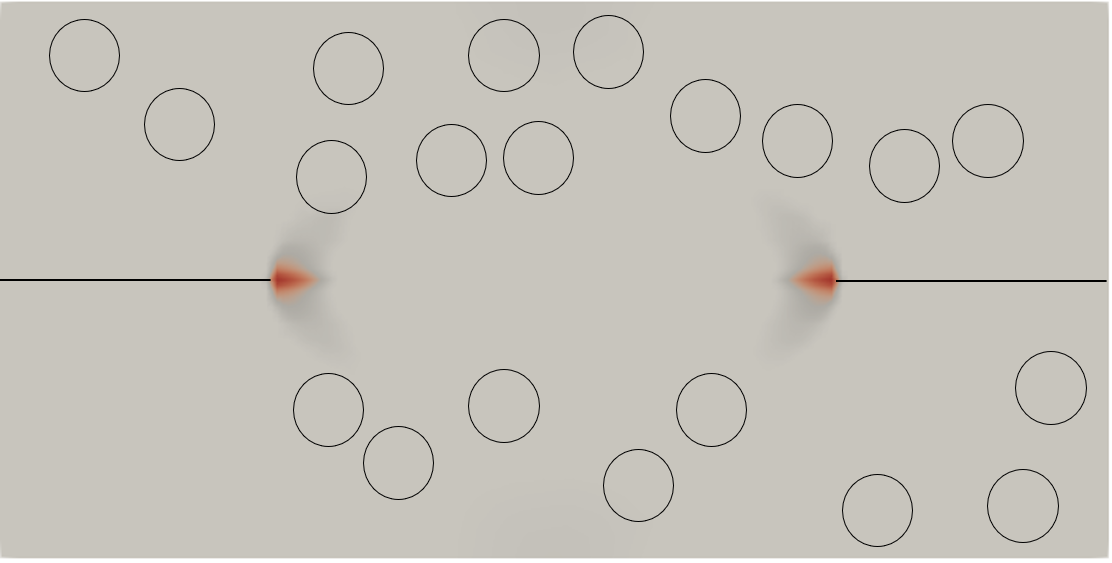}
\caption{}
\end{subfigure}
\hfill
\begin{subfigure}{0.48\textwidth}
\centering
\includegraphics[width=\linewidth]{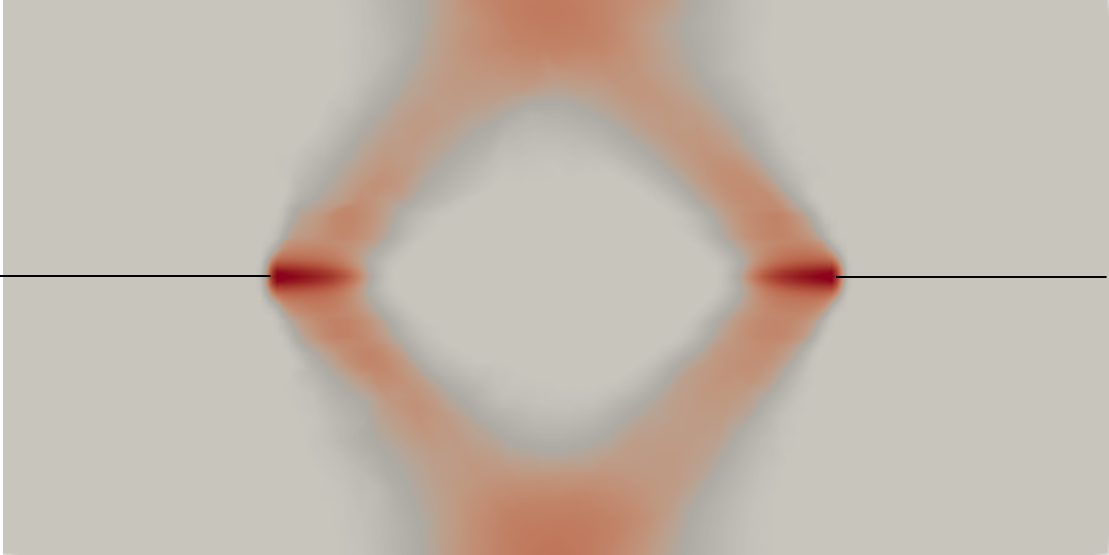}
\caption{}
\end{subfigure}

\vspace{0.2cm}

\begin{subfigure}{0.48\textwidth}
\centering
\includegraphics[width=\linewidth]{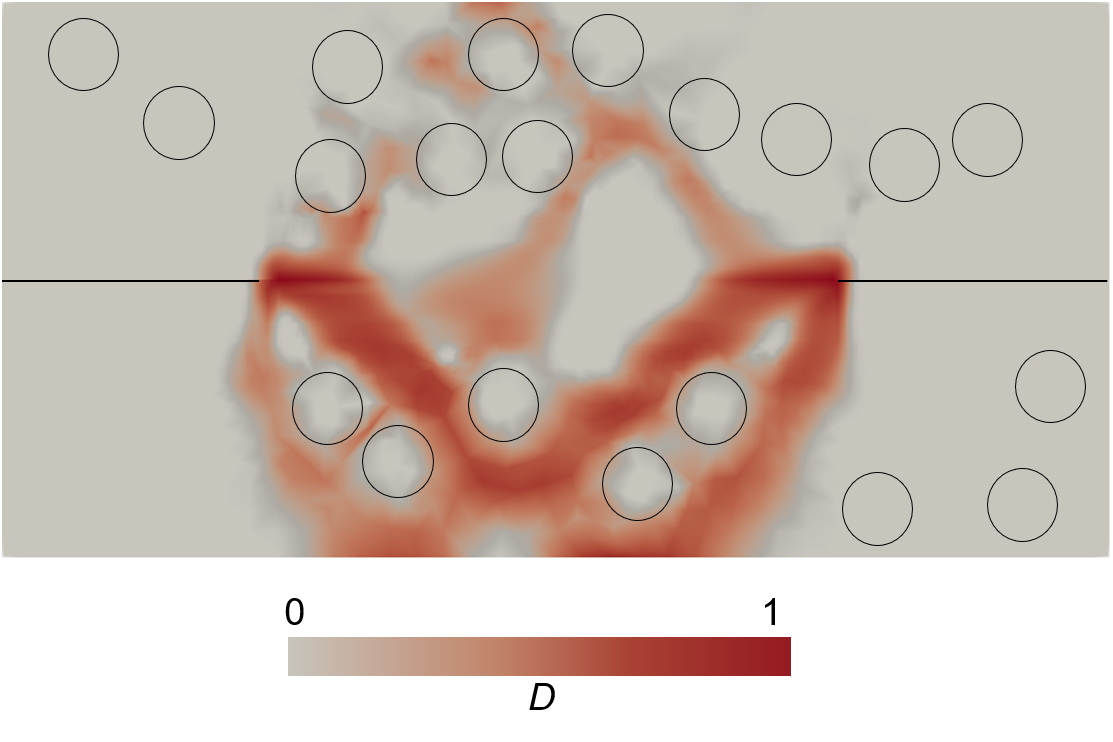}
\caption{}
\end{subfigure}
\hfill
\begin{subfigure}{0.48\textwidth}
\centering
\includegraphics[width=\linewidth]{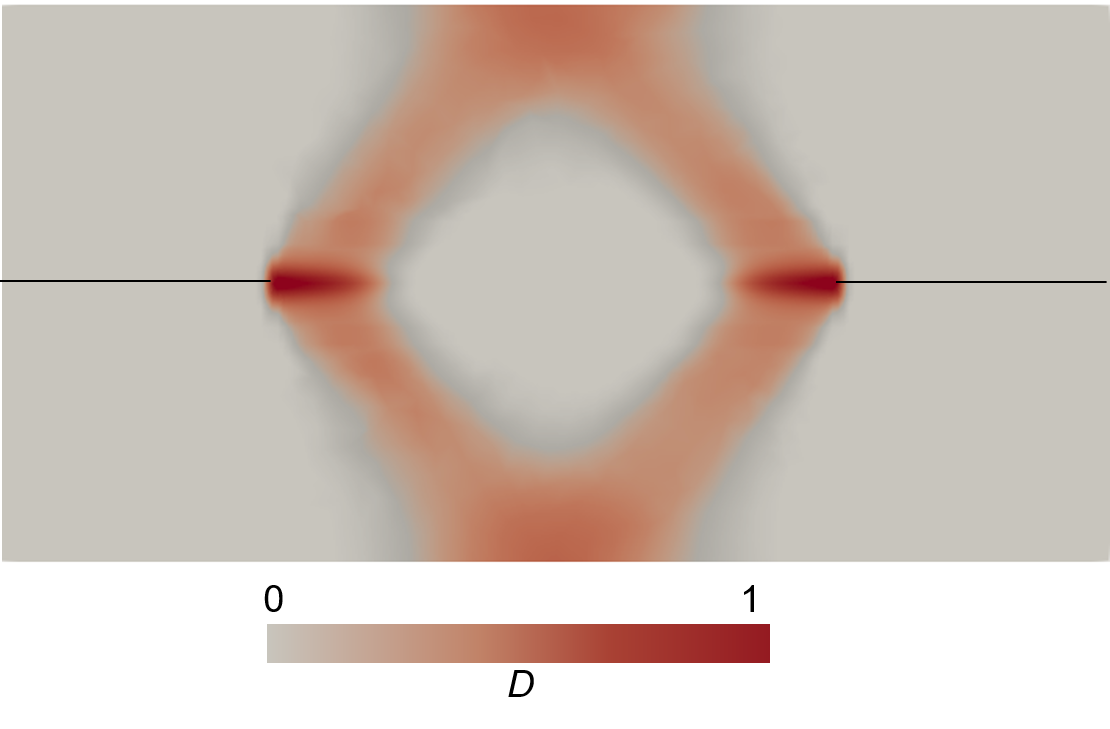}
\caption{}
\end{subfigure}

\caption{Comparison of damage contours obtained from the heterogeneous composite (left column) and the corresponding homogenized model (right column) at displacement levels of $u=0.25~\mathrm{mm}$, $u=0.275~\mathrm{mm}$, and $u=0.40~\mathrm{mm}$.}
\label{fig:damage_5percent}
\end{figure}
Figures~\ref{fig:damage_5percent}(a,b), \ref{fig:damage_5percent}(c,d), and \ref{fig:damage_5percent}(e,f) compare the damage evolution in the heterogeneous composite and the corresponding homogenized model at different stages of loading. Although the specimen contains a symmetric double-edge crack configuration, damage propagation in the heterogeneous composite deviates noticeably from symmetry. This behavior arises from the random distribution and orientation of the inhomogeneities, which locally alter the stress field and influence the preferred damage path. It can further be observed that the damage does not propagate through the inhomogeneities; instead, the propagation path is deflected when interacting with the reinforcing particles, resulting in a tortuous damage trajectory. In contrast, the homogenized model exhibits a more symmetric and smoother damage evolution pattern, since the influence of the discrete inhomogeneities is represented through the effective constitutive response rather than explicit microstructural interactions. Nevertheless, the overall damage progression and the location of the dominant damaged region remain consistent between the heterogeneous and homogenized representations.
\subsubsection{Composite with 15\% and 25\% Volume Fractions}
In the previous subsection, a composite containing a $5\%$ inclusion volume fraction was considered, where the particles occupied a relatively small portion of the domain and were therefore well separated from one another. To further assess the robustness and predictive capability of the proposed framework, composites with inclusion volume fractions of $15\%$ and $25\%$ are subsequently investigated. For consistency, the total number of particles is maintained constant across all three cases. Consequently, the particle radius increases with increasing volume fraction, resulting in reduced inter-particle spacing and enhanced particle interactions. Furthermore, a new random particle distribution is generated for each volume fraction to ensure a statistically representative microstructure while satisfying the prescribed inclusion content.
\begin{figure}[H]
    \centering

    \begin{subfigure}{0.48\textwidth}
        \centering
        \includegraphics[width=\linewidth]{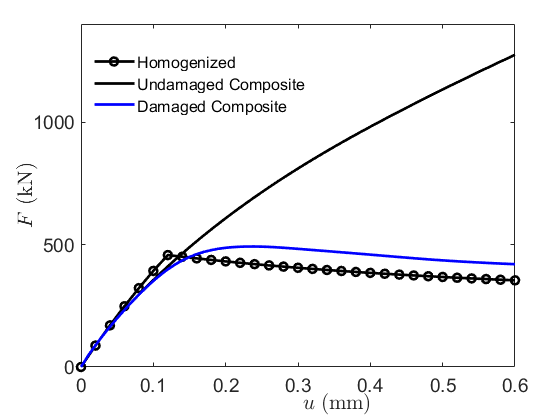}
        \caption{}
        \label{fig15a}
    \end{subfigure}
    \hfill
    \begin{subfigure}{0.48\textwidth}
        \centering
        \includegraphics[width=\linewidth]{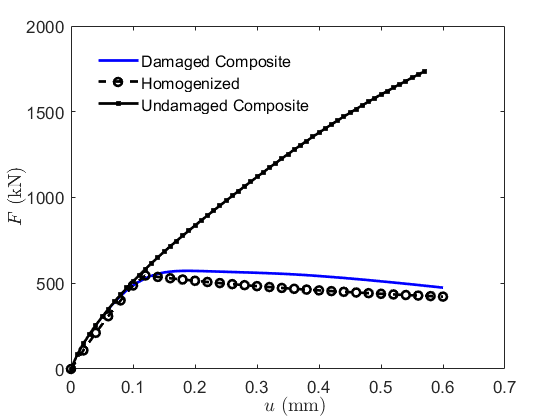}
        \caption{}
        \label{fig15b}
    \end{subfigure}

    \caption{Force-displacement responses of the undamaged composite, damaged heterogeneous composite, and corresponding homogenized model for inclusion volume fractions of (a) $15\%$ and (b) $25\%$.}
    \label{fig:fd_15_25}
\end{figure}
A clear increase in the load-carrying capacity of the structure can be observed with increasing inclusion volume fraction. The peak load increases from approximately $410~\mathrm{kN}$ for the $5\%$ volume fraction case to about $585~\mathrm{kN}$ for the $25\%$ volume fraction case, demonstrating the enhanced reinforcing effect provided by the inhomogeneities. Since the damage evolution characteristics and propagation patterns remain qualitatively similar to those discussed for the $5\%$ volume fraction, the damage contour plots are omitted here to avoid redundancy.

\begin{table}[H]
\centering
\caption{Comparison of mesh characteristics and computational cost for the heterogeneous and homogenized models.}
\label{tab:cost}
\begin{tabular}{ccccccc}
\hline
Volume &
\multicolumn{3}{c}{Heterogeneous Model} &
\multicolumn{3}{c}{Homogenized Model}\\
\cline{2-4}\cline{5-7}
Fraction &
Mesh Size &
Elements &
CPU Time &
Mesh Size &
Elements &
CPU Time\\
(\%) & (mm) & (-) & (s) & (mm) & (-) & (s)\\
\hline
5  & 0.025  & 3200  & 10800  &0.025  &3200  &7717  \\
15 & 0.0224 & 4000  &12600  &0.025  &3200  &7782  \\
25 & 0.0180  & 6200 & 18000 &0.025  & 3200 &7705  \\
\hline
\end{tabular}
\end{table}
The computational statistics reported in this study were obtained on a workstation equipped with an Intel\textsuperscript{\textregistered} Core\textsuperscript{\texttrademark} i7-10700F processor operating at 2.90 GHz, comprising 8 physical cores and 16 logical processors, together with 64 GB of RAM. All finite element simulations are performed using in-house MATLAB codes developed by the authors. It should be noted that no parallelization techniques or multi-core processing strategies are employed; therefore, the reported computational times correspond to serial execution.

The effective element size reported in Table~\ref{tab:cost} is estimated based on the characteristic element length associated with the finite element discretization. This measure is particularly important for accurately capturing strain-softening and damage localization phenomena and provides a consistent basis for comparing meshes with different levels of refinement. The estimation procedure follows the methodology proposed by Pabbu \textit{et al.}~\cite{Pabbu2025}.

\section{Summary and Conclusions}\label{S4}
In the present work, a second-order homogenization framework is successfully integrated with structural-scale finite element analyses of particulate elastomeric composites. The methodology is validated against benchmark results from the literature and subsequently employed to identify equivalent hyperelastic material parameters from the effective nominal stress-stretch response. The resulting homogenized models showed the ability to reproduce the structural response of the heterogeneous composites for both two- and three-dimensional problems. Furthermore, it is observed that the isotropic approximation of Hill's polarization tensor exhibits numerical difficulties for nearly incompressible case, whereas the integral expression of the polarization tensor provided stable predictions.

The effective properties evaluated from the homogenization framework are later used with a mixed displacement-pressure finite element formulation and an isotropic continuum damage model to investigate matrix-dominated damage evolution. A scaling procedure is proposed to determine an equivalent damage initiation threshold for the homogenized domain, enabling the homogenized models to capture the softening response of the heterogeneous composites with good accuracy. The numerical investigations demonstrate that the proposed homogenization framework closely reproduces the global force--displacement response and the overall damage-induced stiffness degradation of the heterogeneous composite, while offering a significant reduction in computational cost. However, computational efficiency is achieved by replacing the heterogeneous microstructure with an equivalent homogeneous constitutive description; consequently, local microscale damage mechanisms are not explicitly resolved.


\section*{Acknowledgments}
 This work was supported by DST-SERB and VSSC, and ISRO of the project titled "Functionality Enhancement through Design and Development of Advanced Finite Element Algorithms for STR Tools" under the IMPRINT.IIC scheme (IMP/2019/000276) and Anusandhan National Research Foundation (ANRF) (formerly DST-SERB) PMECRG grant (Project File No. ANRF/ECRG/2025/000032/ENS).
 
  \section*{Data Availability Statement}
Data sharing not applicable to this article as no datasets were generated or analyzed during the current study.
\appendix
\renewcommand{\theequation}{\Alph{section}.\arabic{equation}}
\section{Second-Order Homogenization Method}\label{Appenidx A}
\setcounter{equation}{0}
The following assumptions are made in deriving the final expressions for
$\overline{\psi}$ and $\overline{\boldsymbol{P}}$:
\begin{itemize}
    \item[\textbf{(A1)}] The functions
    $\psi_{\gamma}$ are sufficiently smooth.

    \item[\textbf{(A2)}] The phase strain-energy
functions $\psi_{\gamma}$ are assumed to satisfy polyconvexity with respect
to $\boldsymbol{F}$. However, the energy
difference $(\psi-\psi_{\mathrm{o}})$ does not necessarily retain the
convexity required for an extremum-based variational formulation.
Therefore, the extremum operation is replaced by a stationary operation,
which provides estimates of the effective response rather than variational
bounds.

    \item[\textbf{(A3)}] Each phase satisfies the impenetrability condition,
    i.e., $\det\boldsymbol{F}(\boldsymbol{X})>0$.
\end{itemize}

The energy difference $\Delta\psi$ introduced in the main text (Eq.~\ref{eq10}) is used here to derive its complementary representation through the Legendre transform.
The complementary potential is defined as
\begin{equation}\label{eqA1}
\Delta\psi^{*}(\boldsymbol{X},\boldsymbol{T})
=
\operatorname*{stat}_{\boldsymbol{F}}
\left\{
\boldsymbol{T}\cdot\boldsymbol{F}
-
\Delta\psi(\boldsymbol{X},\boldsymbol{F})
\right\}.
\end{equation}
Due to (A1), the inverse Legendre transformation gives
\begin{equation}
\Delta\psi(\boldsymbol{X},\boldsymbol{F})
=
\operatorname*{stat}_{\boldsymbol{T}}
\left\{
\boldsymbol{T}\cdot\boldsymbol{F}
-
\Delta\psi^{*}(\boldsymbol{X},\boldsymbol{T})
\right\}.
\end{equation}
The inverse Legendre transformation gives the local strain-energy density in terms of the deformation gradient and the complementary potential as
\begin{equation}
\psi(\boldsymbol{X},\boldsymbol{F})
=
\operatorname*{stat}_{\boldsymbol{T}}
\left\{
\boldsymbol{T}\cdot\boldsymbol{F}
+
\psi_{\mathrm{o}}(\boldsymbol{X},\boldsymbol{F})
-
\Delta\psi^{*}(\boldsymbol{X},\boldsymbol{T})
\right\},
\label{eq:app_inverse_legendre}
\end{equation}
where $\Delta\psi^{*}=(\psi-\psi_{\mathrm{o}})^{*}$ denotes the complementary potential of the energy difference.

Substitution of Eq.~(\ref{eq:app_inverse_legendre}) into Eq.~(\ref{eq6}) gives
\begin{equation}
\overline{\psi}(\overline{\boldsymbol{F}})
=
\operatorname*{stat}_{\boldsymbol{F}\in\mathcal{K}(\overline{\boldsymbol{F}})}
\sum_{\gamma=1}^{N}
c_{\gamma}
\Bigg\langle
\operatorname*{stat}_{\boldsymbol{T}}
\left\{
\boldsymbol{T}\cdot\boldsymbol{F}
+
\psi_{\mathrm{o}}(\boldsymbol{X},\boldsymbol{F})
-
\Delta\psi^{*}(\boldsymbol{X},\boldsymbol{T})
\right\}
\Bigg\rangle .
\label{eq:app_substitution}
\end{equation}

Because the assumption (A1), the two stationary operations in Eq.~(\ref{eq:app_substitution}) may be
interchanged. Since $\Delta\psi^{*}(\boldsymbol{X},\boldsymbol{T})$ is
independent of $\boldsymbol{F}$, it is unaffected by the stationary
operation with respect to $\boldsymbol{F}$ and can therefore be taken
outside the inner stationary operation. In addition, the linearity of the
volume-average operator permits the averaging of the individual terms
separately. Thus,
\begin{equation}
\overline{\psi}(\overline{\boldsymbol{F}})
=
\operatorname*{stat}_{\boldsymbol{T}}
\Bigg\{
\sum_{\gamma=1}^{N}
c_{\gamma}
\operatorname*{stat}_{\boldsymbol{F}\in\mathcal{K}(\overline{\boldsymbol{F}})}
\left\langle
\boldsymbol{T}\cdot\boldsymbol{F}
+
\psi_{\mathrm{o}}(\boldsymbol{X},\boldsymbol{F})
\right\rangle
-
\left\langle
\Delta\psi^{*}(\boldsymbol{X},\boldsymbol{T})
\right\rangle
\Bigg\}.
\label{eq:app_final_variational}
\end{equation}

The polarization field is represented in a phase-wise form using the
indicator function $\mathcal{J}_{\gamma}(\boldsymbol{X})$ of phase $\gamma$
as
\begin{equation}
\boldsymbol{T}(\boldsymbol{X})
=
\sum_{\gamma=1}^{N}
\mathcal{J}_{\gamma}(\boldsymbol{X})\,
\boldsymbol{T}_{\gamma},
\end{equation}
where $\boldsymbol{T}_{\gamma}$ denotes the polarization tensor associated
with phase $\gamma$. Substituting this representation into the stationary
variational expression and using the phase-wise form of the complementary
potential gives
\begin{equation}\label{eqA8}
\begin{aligned}
\overline{\psi}(\overline{\boldsymbol{F}})
&=
\operatorname*{stat}_{\boldsymbol{T}}
\Bigg\{
\operatorname*{stat}_{\boldsymbol{F}\in\mathcal{K}(\overline{\boldsymbol{F}})}
\left\langle
\boldsymbol{T}\cdot\boldsymbol{F}
+
\psi_{\mathrm{o}}(\boldsymbol{X},\boldsymbol{F})
\right\rangle
-
\left\langle
(\psi-\psi_{\mathrm{o}})^{*}
(\boldsymbol{X},\boldsymbol{T})
\right\rangle
\Bigg\}
\\[3mm]
&=
\operatorname*{stat}_{\boldsymbol{T}}
\Bigg\{
\operatorname*{stat}_{\boldsymbol{F}\in\mathcal{K}(\overline{\boldsymbol{F}})}
\left\langle
\boldsymbol{T}\cdot\boldsymbol{F}
+
\psi_{\mathrm{o}}(\boldsymbol{X},\boldsymbol{F})
\right\rangle
-
\sum_{\gamma=1}^{N}
c_{\gamma}
\left(
\psi_{\gamma}
-
\psi_{\mathrm{o}_{\gamma}}
\right)^{*}
\!\left(
\boldsymbol{T}_{\gamma}
\right)
\Bigg\}.
\end{aligned}
\end{equation}
The stationary contribution with respect to the deformation field is
defined as
\begin{equation}\label{eqA8}
\overline{\psi}_{\uptheta}
\left(
\overline{\boldsymbol{F}};
\boldsymbol{T}
\right)
=
\operatorname*{stat}_{\boldsymbol{F}\in\mathcal{K}(\overline{\boldsymbol{F}})}
\Big\langle
\boldsymbol{T}\cdot\boldsymbol{F}
+
\psi_{\mathrm{o}}(\boldsymbol{X},\boldsymbol{F})
\Big\rangle.
\end{equation}
Equation~(\ref{eqA8}) represents the effective free energy of the fictitious linear thermoelastic composite. The strong-form equations (equilibrium) and boundary conditions yield
\begin{equation}
\left\langle\boldsymbol{F}\right\rangle
=
\overline{\boldsymbol{F}},
\end{equation}
which, upon decomposition into phase averages, gives
\begin{equation}
\sum_{\gamma=1}^{N}
c_{\gamma}\overline{\boldsymbol{F}}_{\gamma}
=
\overline{\boldsymbol{F}}.
\end{equation}

The stationarity condition of Eq.~(\ref{eqA1}) relates the polarization tensor to the difference between the nominal stresses of the actual material and the LCC, which is given as,
\begin{align}
\boldsymbol{T}
&=
\frac{\partial(\psi-\psi_{\mathrm{o}})}
{\partial\boldsymbol{F}}
(\boldsymbol{X},\boldsymbol{F})
\nonumber\\
&=
\frac{\partial\psi(\boldsymbol{X},\boldsymbol{F})}
{\partial\boldsymbol{F}}
-
\frac{\partial\psi_{\mathrm{o}}(\boldsymbol{X},\boldsymbol{F})}
{\partial\boldsymbol{F}}
\nonumber\\
&=
\boldsymbol{P}
-
\mathbb{A}_{\mathrm{o}}[\boldsymbol{F}].
\label{eq:T_relation}
\end{align}
The relation is obtained using Eq.~(\ref{eq:P_definition}) and the quadratic form of the LCC strain-energy density function.
For a heterogeneous composite with phase-wise comparison properties, the relation for phase $\gamma$ is therefore
\begin{equation}
\boldsymbol{T}_{\gamma}
=
\boldsymbol{P}_{\gamma}
-
\mathbb{A}_{\mathrm{o}_{\gamma}}
[\boldsymbol{F}_{\gamma}].
\label{eq:T_gamma_relation}
\end{equation}
Here, $\boldsymbol{F}_{\gamma}$ is dual to $\boldsymbol{T}_{\gamma}$. The phase relation is analogous to the constitutive relation of linear thermoelasticity in a small-strain setting. Substituting the phase-wise relation for the polarization tensor into the stationary representation of the effective strain-energy density gives
\begin{align}\label{eqA13}
\overline{\psi}(\overline{\boldsymbol{F}})
&=
\operatorname*{stat}_{\boldsymbol{T}}
\Bigg\{
\overline{\psi}_{\uptheta}
(\overline{\boldsymbol{F}};\boldsymbol{T})
-
\sum_{\gamma=1}^{N}
c_\gamma
\left(
\psi_\gamma
-
\psi_{\mathrm{o}_\gamma}
\right)^{*}
\!\left(
\boldsymbol{T}_\gamma
\right)
\Bigg\}.
\end{align}
Using the quadratic form of the LCC strain-energy density and Eq.~(\ref{eq:T_gamma_relation}),
the first contribution in the above expression can be written as
\begin{equation}
\begin{aligned}
\overline{\psi}_{\uptheta}
(\overline{\boldsymbol{F}};\boldsymbol{T})
&=
\operatorname*{stat}_{\boldsymbol{F}\in\mathcal{K}(\overline{\boldsymbol{F}})}
\sum_{\gamma=1}^{N}
c_{\gamma}
\left\langle
\left(
\boldsymbol{P}_\gamma
-
\mathbb{A}_{\mathrm{o}_{\gamma}}
[\boldsymbol{F}_{\gamma}]
\right)\cdot\boldsymbol{F}
+
\frac{1}{2}
\boldsymbol{F}\cdot
\mathbb{A}_{\mathrm{o}}[\boldsymbol{F}]
\right\rangle_{\gamma}.
\end{aligned}
\label{eqA15}
\end{equation}

At this point, it is to be noted that $\boldsymbol{F}$ denotes the deformation gradient related to the thermoelasticity problem, which is inhomogeneous. The second contribution is given by

\begin{equation}
\left(
\psi_\gamma
-
\psi_{\mathrm{o}_\gamma}
\right)^{*}
\left(
\boldsymbol{T}_\gamma
\right)
=
\left(
\boldsymbol{P}_\gamma\cdot
\boldsymbol{F}_\gamma
-
\psi_\gamma(\boldsymbol{F}_\gamma)
-
\frac{1}{2}
\boldsymbol{F}_\gamma\cdot
\mathbb{A}_{\mathrm{o}_\gamma}
[\boldsymbol{F}_\gamma]
\right).
\label{eqA15b}
\end{equation}
The quadratic terms can then be rearranged by introducing the deformation fluctuation
\begin{equation}
\Delta\boldsymbol{F}_\gamma
=
\boldsymbol{F}-\boldsymbol{F}_\gamma,
\end{equation}
Upon substitution of Eq.~(\ref{eqA15}) and Eq.~(\ref{eqA15b}) in Eq.~(\ref{eqA13}) gives
\begin{align*}
\overline{\psi}(\overline{\boldsymbol{F}})
&=
\sum_{\gamma=1}^{N}
c_{\gamma}
\operatorname*{stat}_{\boldsymbol{F}}
\Bigg\{
\psi_\gamma(\boldsymbol{F}_\gamma)
+
\boldsymbol{P}_\gamma\cdot
\left\langle
\Delta\boldsymbol{F}_\gamma
\right\rangle_\gamma
+
\frac{1}{2}
\left\langle
\Delta\boldsymbol{F}_\gamma
\cdot
\mathbb{A}_{\mathrm{o}_\gamma}
[
\Delta\boldsymbol{F}_\gamma]
\right\rangle_\gamma
\Bigg\}.
\end{align*}
The stationary condition with respect to the phase-average deformation gradient
$\boldsymbol{F}_\gamma$ is obtained by requiring
\begin{equation}
\frac{\partial\overline{\psi}}
{\partial\boldsymbol{F}_\gamma}
=
\boldsymbol{0}.
\end{equation}
Using the expression for the effective strain-energy density, this condition becomes
\begin{equation}
\frac{\partial}{\partial\boldsymbol{F}_\gamma}
\Bigg(
\psi_\gamma(\boldsymbol{F}_\gamma)
+
\boldsymbol{P}_\gamma\cdot
\left\langle
\Delta\boldsymbol{F}_\gamma
\right\rangle_\gamma
+
\frac{1}{2}
\left\langle
\Delta\boldsymbol{F}_\gamma
\cdot
\mathbb{A}_{\mathrm{o}_\gamma}
[
\Delta\boldsymbol{F}_\gamma]
\right\rangle_\gamma
\Bigg)
=
\boldsymbol{0}.
\end{equation}
Using
\begin{equation}
\frac{\partial\psi_\gamma}{\partial\boldsymbol{F}_\gamma}
=
\boldsymbol{P}_\gamma,
\qquad
\frac{\partial\boldsymbol{P}_\gamma}
{\partial\boldsymbol{F}_\gamma}
=
\mathbb{A}_{\gamma},
\end{equation}
and
\begin{equation}
\frac{\partial
\left\langle
\Delta\boldsymbol{F}_\gamma
\right\rangle_\gamma}
{\partial\boldsymbol{F}_\gamma}
=
-\mathbb{I},
\end{equation}
the stationarity condition reduces to
\begin{equation}
\boldsymbol{P}_\gamma
+
\mathbb{A}_{\gamma}
[
\left\langle
\Delta\boldsymbol{F}_\gamma
\right\rangle_\gamma]
-
\boldsymbol{P}_\gamma
-
\mathbb{A}_{\mathrm{o}_\gamma}
[
\left\langle
\Delta\boldsymbol{F}_\gamma
\right\rangle_\gamma]
=
\boldsymbol{0}.
\end{equation}
Hence,
\begin{equation}\label{A22}
c_\gamma\left(
\mathbb{A}_{\gamma}
-
\mathbb{A}_{\mathrm{o}_\gamma}
\right)
\left[
\left\langle
\Delta\boldsymbol{F}_\gamma
\right\rangle_\gamma\right]
=
\boldsymbol{0}.
\end{equation}
Using Eq.(\ref{A22}) deduces to  
\begin{equation*}
\mathbb{A}_{\gamma}
=
\mathbb{A}_{\mathrm{o}_\gamma},\quad
\langle
\boldsymbol{F}\rangle_\gamma=\overline{\boldsymbol{F}}_\gamma=
\boldsymbol{F}_\gamma
\end{equation*}
Using the stationarity conditions derived above together with the equilibrium
condition of the LCC, the $\overline{\psi}$ is expressed as
\begin{align}
\overline{\psi}(\overline{\boldsymbol{F}})
=\sum_{\gamma=1}^{N}c_\gamma\left(
\psi_\gamma\!\left(\overline{\boldsymbol{F}}_\gamma\right)
+
\frac{1}{2}
\left(
\overline{\boldsymbol{F}}
-
\overline{\boldsymbol{F}}_\gamma
\right)
\cdot
\frac{\partial \psi_\gamma
\!\left(\overline{\boldsymbol{F}}_\gamma\right)}
{\partial \boldsymbol{F}}\right).
\label{eq:app_effpsi1}
\end{align}
\begin{align}
\overline{\psi}(\overline{\boldsymbol{F}})
=\sum_{\gamma=1}^{N}c_\gamma\left(
\psi_\gamma\!\left(\overline{\boldsymbol{F}}_\gamma\right)
+
\frac{1}{2}
\left(
\overline{\boldsymbol{F}}
-
\overline{\boldsymbol{F}}_\gamma
\right)
\cdot
\boldsymbol{P}_\gamma
\!\left(
\overline{\boldsymbol{F}}_\gamma
\right)\right).
\label{eq:app_effpsi2}
\end{align}
The $\overline{\boldsymbol{P}}$ is obtained by differentiating the $\overline{\psi}$ with respect to the $\overline{\boldsymbol{F}}$
The first term in Eq.~(\ref{eq:app_effpsi2}) depends on
$\overline{\boldsymbol{F}}$ through
$\overline{\boldsymbol{F}}_\gamma$. Therefore, applying the chain rule gives
\begin{align}
\frac{\partial}
{\partial\overline{\boldsymbol{F}}}
\psi_\gamma
\!\left(
\overline{\boldsymbol{F}}_\gamma
\right)
&=
\frac{\partial
\psi_\gamma
\!\left(
\overline{\boldsymbol{F}}_\gamma
\right)}
{\partial\overline{\boldsymbol{F}}_\gamma}
\cdot
\frac{\partial
\overline{\boldsymbol{F}}_\gamma}
{\partial\overline{\boldsymbol{F}}}
\nonumber\\
&=
\left(
\frac{\partial
\overline{\boldsymbol{F}}_\gamma}
{\partial\overline{\boldsymbol{F}}}\right)^{\mathrm{T}}\left[
\boldsymbol{P}_\gamma
\!\left(
\overline{\boldsymbol{F}}_\gamma
\right)\right].
\label{eq:app_chain}
\end{align}
The second term in Eq.~(\ref{eq:app_effpsi2}) contains both
$\overline{\boldsymbol{F}}$ and
$\overline{\boldsymbol{F}}_\gamma$, and is therefore differentiated using
the product rule. It is first separated as
\begin{align}
&\frac{\partial}
{\partial\overline{\boldsymbol{F}}}
\Bigg[
\frac{1}{2}
\left(
\overline{\boldsymbol{F}}
-
\overline{\boldsymbol{F}}_\gamma
\right)
\cdot
\boldsymbol{P}_\gamma
\Bigg]
\nonumber\\
&=
\frac{1}{2}
\Bigg[
\frac{\partial}
{\partial\overline{\boldsymbol{F}}}
\left(
\overline{\boldsymbol{F}}
\cdot
\boldsymbol{P}_\gamma
\right)
-
\frac{\partial}
{\partial\overline{\boldsymbol{F}}}
\left(
\overline{\boldsymbol{F}}_\gamma
\cdot
\boldsymbol{P}_\gamma
\right)
\Bigg].
\label{eq:app_product}
\end{align}
Using
\[
\frac{\partial\boldsymbol{P}_\gamma}
{\partial\overline{\boldsymbol{F}}_\gamma}
=
\mathbb{A}_{\gamma},
\]
$({(A.26)_1})$ is evaluated as

In index notation,
\begin{align}
\frac{\partial
\left(
\overline{F}_{ij}P_{ij}^{\gamma}
\right)}
{\partial\overline{F}_{kl}}
&=
\frac{\partial \overline{F}_{ij}}
{\partial\overline{F}_{kl}}
P_{ij}^{\gamma}
+
\overline{F}_{ij}
\frac{\partial P_{ij}^{\gamma}}
{\partial\overline{F}_{kl}}=
\frac{\partial \overline{F}_{ij}}
{\partial\overline{F}_{kl}}
P_{ij}^{\gamma}
+
\overline{F}_{ij}\frac{\partial P_{ij}^{\gamma}}
{\partial{\overline{F}}^\gamma_{mn}}
\frac{\partial{\overline{F}}^\gamma_{mn}}
{\partial\overline{F}_{kl}}.
\label{eq:app_product1_in}
\end{align}
\begin{align}
\frac{\partial}
{\partial\overline{\boldsymbol{F}}}
\left(
\overline{\boldsymbol{F}}
\cdot
\boldsymbol{P}_\gamma
\right)
&=
\boldsymbol{P}_\gamma
+
\left(
\mathbb{A}_{\gamma}
\left(
\frac{\partial
\overline{\boldsymbol{F}}_\gamma}
{\partial\overline{\boldsymbol{F}}}\right)\right)^{\mathrm{T}}\left[\overline{\boldsymbol{F}}\right].
\label{eq:app_product1}
\end{align}

$({(A.26)_2})$ gives
\begin{align}
\frac{\partial}
{\partial\overline{\boldsymbol{F}}}
\left(
\overline{\boldsymbol{F}}_\gamma
\cdot
\boldsymbol{P}_\gamma
\right)
&=
\left(
\frac{\partial
\overline{\boldsymbol{F}}_\gamma}
{\partial\overline{\boldsymbol{F}}}
\right)^{\mathrm{T}}
[\boldsymbol{P}_\gamma]
+
\left(
\mathbb{A}_{\gamma}
\left(
\frac{\partial
\overline{\boldsymbol{F}}_\gamma}
{\partial\overline{\boldsymbol{F}}}\right)\right)^{\mathrm{T}}\left[\overline{\boldsymbol{F}}_\gamma\right].
\label{eq:app_product2}
\end{align}
Substituting Eqs.~(\ref{eq:app_product1}) and
(\ref{eq:app_product2}) into Eq.~(\ref{eq:app_product}) gives
\begin{align}
&\frac{\partial}
{\partial\overline{\boldsymbol{F}}}
\Bigg[
\frac{1}{2}
\left(
\overline{\boldsymbol{F}}
-
\overline{\boldsymbol{F}}_\gamma
\right)
\cdot
\boldsymbol{P}_\gamma
\Bigg]
\nonumber\\
&=
\frac{1}{2}
\Bigg(
\boldsymbol{P}_\gamma
+
\left(
\mathbb{A}_{\gamma}
\left(
\frac{\partial
\overline{\boldsymbol{F}}_\gamma}
{\partial\overline{\boldsymbol{F}}}\right)\right)^{\mathrm{T}}
[\overline{\boldsymbol{F}}]
-
\left(
\frac{\partial
\overline{\boldsymbol{F}}_\gamma}
{\partial\overline{\boldsymbol{F}}}\right)^{\mathrm{T}}
[\boldsymbol{P}_\gamma]
-
\left(
\mathbb{A}_{\gamma}
\left(
\frac{\partial
\overline{\boldsymbol{F}}_\gamma}
{\partial\overline{\boldsymbol{F}}}\right)\right)^{\mathrm{T}}
\left[\overline{\boldsymbol{F}}_\gamma\right]
\Bigg).
\label{eq:app_second_derivative}
\end{align}
Combining the derivative of  $({(A.26)_1})$ with the derivative of $({(A.26)_2})$ gives
\begin{equation}
\begin{aligned}
\boldsymbol{\overline{P}}
&=
\sum_{\gamma=1}^{N}c_\gamma
\left[
\begin{aligned}
&\left(
\frac{\partial
\overline{\boldsymbol{F}}_\gamma}
{\partial\overline{\boldsymbol{F}}}\right)^{\mathrm{T}}
\left[\boldsymbol{P}_\gamma\right]
\\[2mm]
&\quad+
\frac{1}{2}
\Bigg(
\boldsymbol{P}_\gamma
+
\left(
\mathbb{A}_{\gamma}
\left(
\frac{\partial
\overline{\boldsymbol{F}}_\gamma}
{\partial\overline{\boldsymbol{F}}}
\right)\right)^{\mathrm{T}}
[\overline{\boldsymbol{F}}]
-
\left(
\frac{\partial
\overline{\boldsymbol{F}}_\gamma}
{\partial\overline{\boldsymbol{F}}}\right)^{\mathrm{T}}
[\boldsymbol{P}_\gamma]
-
\left(
\mathbb{A}_{\gamma}
\left(
\frac{\partial
\overline{\boldsymbol{F}}_\gamma}
{\partial\overline{\boldsymbol{F}}}
\right)\right)^{\mathrm{T}}
[\overline{\boldsymbol{F}}_\gamma]
\Bigg)
\end{aligned}
\right].
\end{aligned}
\label{eq:app_sigma1}
\end{equation}
Collecting the terms $\left(\frac{\partial\overline{\boldsymbol{F}}_\gamma}
{\partial\overline{\boldsymbol{F}}}\right)$ and factoring the tangent contribution gives
\begin{align}
\boldsymbol{\overline{P}}
&=
\frac{1}{2}\sum_{\gamma=1}^{N}c_\gamma
\Bigg(
2\left(
\frac{\partial
\overline{\boldsymbol{F}}_\gamma}
{\partial\overline{\boldsymbol{F}}}\right)^{\mathrm{T}}
[\boldsymbol{P}_\gamma]
+
\boldsymbol{P}_\gamma
+
\left(
\mathbb{A}_{\gamma}
\left(
\frac{\partial
\overline{\boldsymbol{F}}_\gamma}
{\partial\overline{\boldsymbol{F}}}\right)\right)^{\mathrm{T}}
[\overline{\boldsymbol{F}}]
\nonumber\\
&\qquad
-
\left(
\frac{\partial
\overline{\boldsymbol{F}}_\gamma}
{\partial\overline{\boldsymbol{F}}}\right)^{\mathrm{T}}
[\boldsymbol{P}_\gamma]
-
\left(
\mathbb{A}_{\gamma}
\left(
\frac{\partial
\overline{\boldsymbol{F}}_\gamma}
{\partial\overline{\boldsymbol{F}}}\right)\right)^{\mathrm{T}}
[\overline{\boldsymbol{F}}_\gamma]
\Bigg).
\label{eq:app_sigma2}
\end{align}
Rearranging the terms containing
$\mathbb{A}_{\gamma}$ and
$\overline{\boldsymbol{F}}-\overline{\boldsymbol{F}}_\gamma$ yields
\begin{align}
\boldsymbol{\overline{P}}
&=
\frac{1}{2}\sum_{\gamma=1}^{N}c_\gamma
\Bigg(
\boldsymbol{P}_\gamma
+
\left(\frac{\partial
\overline{\boldsymbol{F}}_\gamma}
{\partial\overline{\boldsymbol{F}}}\right)^\mathrm{T} \Big[
\boldsymbol{P}_\gamma
+
\mathbb{A}_{\gamma}
\left[
\overline{\boldsymbol{F}}
-
\overline{\boldsymbol{F}}_\gamma
\right]
\Big]
\Bigg).
\label{eq:app_sigma_final}
\end{align}
\section{Mooney-Rivlin Material Model}\label{Appendix B}
\setcounter{equation}{0}
	The strain energy density potential function $(\psi)$ is given as:
	\begin{equation*}
		\psi=C_{10}(\bar{I_{1}}-3)+C_{01}(\bar{I_{2}}-3)+\frac{1}{D_{1}}(J-1)^2
	\end{equation*} 
where, $\bar{I_{1}}=I_{1}J^{\frac{-2}{3}}, \bar{I_{2}}=I_{2}J^{\frac{-4}{3}}, J=\sqrt{\text{det}(C)}$ and $I_{1},I_{2}$ are the invariants of $\boldsymbol{C}$. The following relations are useful while deriving the expressions
\begin{align*}
	\frac{\partial I_{1}}{\partial F_{ij}} &= 2F_{ij}\\
	\frac{\partial I_{2}}{\partial F_{ij}} & =2(I_{1}F_{ij}-F_{ik}C_{jk})\\
	\frac{\partial C_{ij}}{\partial F_{kl}} & = \frac{\partial F_{mi}F_{jm}}{\partial F_{kl}}\\
	&=\frac{\partial F_{mi}}{\partial F_{kl}}F_{jm}+F_{mi}\frac{\partial F_{mj}}{\partial F_{kl}}\\
	&= \delta_{mk}\delta_{il}F_{mj}+F_{mi}\delta_{mk}\delta_{jl}\\
	&=\delta_{il}F_{kj}+F_{ki}\delta_{jl}
\end{align*}
The nominal stress for a hyperelastic material is obtained from Eq.~(\ref{eq:P_definition}) using the chain rule as
\begin{equation}\label{eq:app_P}
\begin{aligned}
\boldsymbol{P}
&=
\frac{\partial\psi}{\partial\boldsymbol{F}}\\
&=
\frac{\partial\psi}{\partial\bar{I}_{1}}
\frac{\partial\bar{I}_{1}}{\partial\boldsymbol{F}}
+
\frac{\partial\psi}{\partial\bar{I}_{2}}
\frac{\partial\bar{I}_{2}}{\partial\boldsymbol{F}}
+
\frac{\partial\psi}{\partial J}
\frac{\partial J}{\partial\boldsymbol{F}}.
\end{aligned}
\end{equation}
Sorting $({(B.1)_1})$ in index notation as
\begin{align*}
\frac{\partial \bar{I_{1}}}{\partial F_{ij}} &=\frac{\partial (I_{1}J^{\frac{-2}{3}})}{\partial F_{ij}}\\
&=\frac{\partial I_{1}}{\partial F_{ij}}J^{\frac{-2}{3}}+I_{1}\frac{\partial J^{\frac{-2}{3}}}{\partial F_{ij}}\\
&=2F_{ij}J^{\frac{-2}{3}}-\frac{2}{3}I_{1}J^{\frac{-2}{3}}F^{-1}_{ji}
\end{align*}
Sorting $({(B.1)_2})$ in index notation as
\begin{align*}
	\frac{\partial \bar{I_{2}}}{\partial F_{ij}}  &=\frac{\partial (I_{2}J^{\frac{-4}{3}})}{\partial F_{ij}}\\
	&=\frac{\partial I_{2}}{\partial F_{ij}}J^{\frac{-4}{3}}+I_{2}\frac{\partial J^{\frac{-4}{3}}}{\partial F_{ij}}\\
	&=2(I_{1}F_{ij}-F_{ik}C_{jk})J^{\frac{-4}{3}}-\frac{4}{3}I_{2}J^{\frac{-4}{3}}F^{-1}_{ji}\\
	&=2I_{1}F_{ij}J^{\frac{-4}{3}}-2F_{ik}C_{jk}J^{\frac{-4}{3}}-\frac{4}{3}I_{2}J^{\frac{-4}{3}}F^{-1}_{ji}
\end{align*}
Sorting $({(B.1)_3})$ in index notation as
\begin{align*}
	\frac{\partial J}{\partial F_{ij}} = JF^{-1}_{ji}
\end{align*}
The terms $\frac{\partial \psi}{\partial \bar{I_{1}}}=C_{10}, \frac{\partial \psi}{\partial \bar{I_{2}}}=C_{01}, \frac{\partial \psi}{\partial J}=\frac{2}{D_{1}}(J-1)$. The nominal stress tensor for the Mooney-Rivlin material model is
\begin{align*}
	P_{ij} &= C_{10}\left[2F_{ij}J^{\frac{-2}{3}}-\frac{2}{3}I_{1}J^{\frac{-2}{3}}F^{-1}_{ji}\right]+C_{01}\left[2I_{1}F_{ij}J^{\frac{-4}{3}}-2F_{ik}C_{jk}J^{\frac{-4}{3}}-\frac{4}{3}I_{2}J^{\frac{-4}{3}}F^{-1}_{ji}\right] \notag \\
	&+\frac{2}{D_{1}}(J-1)JF^{-1}_{ji} 
\end{align*}
\section*{Nominal Elasticity tensor ($\mathbb{A})$}
The  nominal tangent stiffness tensor as defined in Eq.~(\ref{eq:A_definition})
\begin{equation}
\begin{aligned}
A_{ijkl}
&=
\frac{\partial P_{ij}}{\partial F_{kl}}
\\
&=
\frac{\partial}{\partial F_{kl}}
\bigg(
C_{10}
\left[
2F_{ij}J^{-\frac{2}{3}}
-\frac{2}{3}I_{1}J^{-\frac{2}{3}}F^{-1}_{ji}
\right]
\\
&\qquad\qquad
+
C_{01}
\left[
2I_{1}F_{ij}J^{-\frac{4}{3}}
-2F_{ik}C_{jk}J^{-\frac{4}{3}}
-\frac{4}{3}I_{2}J^{-\frac{4}{3}}F^{-1}_{ji}
\right]
\\
&\qquad\qquad
+
\frac{2}{D_{1}}(J-1)JF^{-1}_{ji}
\bigg).
\end{aligned}
\label{eq:app_A}
\end{equation}
$({(B.2)_1})$ in index notation as
\begin{align*}
	2\frac{\partial (F_{ij}J^{\frac{-2}{3}})}{\partial F_{kl}} &=2\frac{\partial F_{ij}}{\partial F_{kl}}J^{\frac{-2}{3}}-F_{ij}\frac{\partial J^{\frac{-2}{3}}}{\partial F_{kl}}\\
	&=2\delta_{ik}\delta_{jl}J^{\frac{-2}{3}}-\frac{4}{3}J^{\frac{-2}{3}}F_{ij}F^{-1}_{lk}\\
	\frac{\partial (\frac{2}{3}I_{1}J^{\frac{-2}{3}}F^{-1}_{ji})}{\partial F_{kl}} &= \frac{2}{3}\frac{\partial I_{1}}{\partial F_{kl}}J^{\frac{-2}{3}}F^{-1}_{ji}+\frac{2}{3}I_{1}\frac{\partial J^{\frac{-2}{3}}}{\partial F_{kl}}F^{-1}_{ji}+\frac{2}{3}I_{1}J^{\frac{-2}{3}}\frac{\partial F^{-1}_{ji}}{\partial F_{kl}}\\
	&=\frac{4}{3}J^{\frac{-2}{3}}F_{kl}F^{-1}_{ji}-\frac{4}{9}I_{1}J^{\frac{-2}{3}}F^{-1}_{lk}F^{-1}_{ji}-\frac{2}{3}I_{1}J^{\frac{-2}{3}}F^{-1}_{kj}F^{-1}_{li}
\end{align*}
The total expression for differentiating the first term is
\begin{align*}
	\frac{\partial (2F_{ij}J^{\frac{-2}{3}}-\frac{2}{3}J^{\frac{-2}{3}}F^{-1}_{ji})}{\partial F_{kl}} &= C_{10}J^{\frac{-2}{3}}\left[2\delta_{ik}\delta_{jl}-\frac{4}{3}F_{ij}F^{-1}_{lk}-\frac{4}{3}F_{kl}F^{-1}_{ji}+\frac{4}{9}I_{1}F^{-1}_{lk}F^{-1}_{ji}+\frac{2}{3}I_{1}F^{-1}_{kj}F^{-1}_{li}\right]
\end{align*}
$({(B.2)_2})$ in index notation as
\begin{align*}
	2\frac{\partial (I_{1}F_{ij}J^{\frac{-4}{3}})}{\partial F_{kl}} &=2\left(\frac{\partial I_{1}}{\partial F_{kl}}F_{ij}J^{\frac{-4}{3}}+I_{1}\frac{\partial F_{ij}}{\partial F_{kl}}J^{\frac{-4}{3}}+I_{1}F_{ij}\frac{\partial J^{\frac{-4}{3}}}{\partial F_{kl}}\right)\\
	&=2\left(2F_{kl}F_{ij}J^{\frac{-4}{3}}+I_{1}\delta_{ik}\delta_{jl}J^{\frac{-4}{3}}-\frac{4}{3}I_{1}F_{ij}J^{\frac{-4}{3}}F^{-1}_{lk}\right)\\
	& \notag \\	
	2\frac{\partial (F_{im}C_{mj}J^{\frac{-4}{3}})}{\partial F_{kl}}&= 2\left(\frac{\partial F_{im}}{\partial F_{kl}}C_{mj}J^{\frac{-4}{3}}+F_{im}\frac{\partial C_{mj}}{\partial F_{kl}}J^{\frac{-4}{3}}+F_{im}C_{mj}\frac{\partial J^{\frac{-4}{3}}}{\partial F_{kl}}\right)\\
	&=2\left(\delta_{ik}\delta_{ml}C_{mj}J^{\frac{-4}{3}}+F_{im}(\delta_{ml}F_{jk}+F_{lm}\delta_{jk})J^{\frac{-4}{3}}-\frac{4}{3}F_{im}C_{mj}J^{\frac{-4}{3}}F^{-1}_{lk}\right)\\
	&=2\left(\delta_{ik}C_{jl}J^{\frac{-4}{3}}+F_{il}F_{jk}J^{\frac{-4}{3}}+F_{im}F_{lm}\delta_{jk}J^{\frac{-4}{3}}-\frac{4}{3}F_{im}C_{mj}F^{-1}_{lk}J^{\frac{-4}{3}}\right)\\
	& \notag\\
	\frac{4}{3} \frac{\partial (I_{2}J^{\frac{-4}{3}}F^{-1}_{ji})}{\partial F_{kl}} &=\frac{4}{3}\left(\frac{\partial I_{2}}{\partial F_{kl}}J^{\frac{-4}{3}}F^{-1}_{ji}+I_{2}\frac{\partial J^{\frac{-4}{3}}}{\partial F_{kl}}F^{-1}_{ji}+I_{2}J^{\frac{-4}{3}}\frac{\partial F^{-1}_{ji}}{\partial F_{kl}}\right)\\
	&=\frac{4}{3}\left(2I_{1}F_{kl}J^{\frac{-4}{3}}F^{-1}_{ji}-2F_{km}C_{ml}J^{\frac{-4}{3}}F^{-1}_{ji}-\frac{4}{3}I_{2}J^{\frac{-4}{3}}F^{-1}_{lk}F^{-1}_{ji}-I_{2}J^{\frac{-4}{3}}F^{-1}_{kj}F^{-1}_{li}\right)\\
	&=\frac{4}{3}\left(2I_{1}J^{\frac{-4}{3}}F_{kl}F^{-1}_{ji}-2J^{\frac{-4}{3}}F_{km}C_{lm}F^{-1}_{ji}-\frac{4}{3}I_{2}J^{\frac{-4}{3}}F^{-1}_{lk}F^{-1}_{ji}-I_{2}J^{\frac{-4}{3}}F^{-1}_{kj}F^{-1}_{li}\right)
\end{align*}
The final expression for differentiating the second term is
\begin{align*}
	\frac{\partial\left(2I_{1}F_{ij}J^{\frac{-4}{3}}-2F_{im}C_{mj}J^{\frac{-4}{3}}-\frac{4}{3}I_{2}J^{\frac{-4}{3}}F^{-1}_{ji}\right)}{\partial F_{kl}} 
	&= C_{01}J^{\frac{-4}{3}} \bigg[
	2\left(2F_{kl}F_{ij}+I_{1}\delta_{ik}\delta_{jl}-\frac{4}{3}I_{1}F_{ij}F^{-1}_{lk}\right) \\[10pt]
	&\quad -2\left(\delta_{ik}C_{jl}+F_{il}F_{jk}+F_{im}F_{lm}\delta_{jk}-\frac{4}{3}F_{im}C_{mj}F^{-1}_{lk}\right) \\[10pt]
	&\quad -\frac{4}{3} \bigg( 
	2I_{1}F_{kl}F^{-1}_{ji} - 2F_{km}C_{lm}F^{-1}_{ji} 
	\bigg) \\[10pt]
	&\quad -\frac{4}{3} \bigg( 
	-\frac{4}{3}I_{2}F^{-1}_{lk}F^{-1}_{ji} - I_{2}F^{-1}_{kj}F^{-1}_{li}
	\bigg)
	\bigg]
\end{align*}
$({(B.2)_3})$ in index notation as
\begin{align*}
	\frac{\partial ((J-1)J\boldsymbol{F}^{-\text{T}})}{\partial F_{kl}} 
	&=\frac{\partial (J^{2}-J)F^{-1}_{ji}}{\partial F_{kl}}+(J^{2}-J)\frac{\partial F^{-1}_{ji}}{\partial F_{kl}}\\
	&=(2J-1)JF^{-1}_{lk}F^{-1}_{ji}-(J^{2}-J)F^{-1}_{kj}F^{-1}_{li}
\end{align*}
The final expression for differentiating the third term is
\begin{align*}
	\frac{2}{D_{1}}\frac{\partial (J-1)JF^{-1}_{ji}}{\partial F_{kl}}=\frac{2}{D_{1}}\left[(2J-1)JF^{-1}_{lk}F^{-1}_{ji}-(J^{2}-J)F^{-1}_{kj}F^{-1}_{li}\right]
\end{align*}
The nominal elasticity tensor for Mooney-Rivlin material model in terms of indicial notations is given as
\begin{align*}
	A_{ijkl} &= \bigg( C_{10}J^{\frac{-2}{3}} \bigg[
	2\delta_{ik}\delta_{jl} - \frac{4}{3}F_{ij}F^{-1}_{lk} 
	-\frac{4}{3}F_{kl}F^{-1}_{ji}+\frac{4}{9}I_{1}F^{-1}_{lk}F^{-1}_{ji} 
	+ \frac{2}{3}I_{1}F^{-1}_{kj}F^{-1}_{li}
	\bigg] \\[10pt]
	&\quad + C_{01}J^{\frac{-4}{3}} \bigg[
	2\left(2F_{kl}F_{ij}+I_{1}\delta_{ik}\delta_{jl}-\frac{4}{3}I_{1}F_{ij}F^{-1}_{lk}\right) \\[10pt]
	&\quad - 2\left(\delta_{ik}C_{jl}+F_{il}F_{jk}+F_{im}F_{lm}\delta_{jk}-\frac{4}{3}F_{im}C_{mj}F^{-1}_{lk}\right) \\[10pt]
	&\quad - \frac{4}{3} \bigg( 
	2I_{1}F_{kl}F^{-1}_{ji} - 2F_{km}C_{lm}F^{-1}_{ji} 
	\bigg) \\[10pt]
	&\quad - \frac{4}{3} \bigg( 
	-\frac{4}{3}I_{2}F^{-1}_{lk}F^{-1}_{ji} - I_{2}F^{-1}_{kj}F^{-1}_{li}
	\bigg)
	\bigg] \\[10pt]
	&\quad + \frac{2}{D_{1}} \bigg[
	(2J-1)JF^{-1}_{lk}F^{-1}_{ji} - (J^{2}-J)F^{-1}_{kj}F^{-1}_{li}
	\bigg] \bigg) 
\end{align*}
 \bibliographystyle{elsarticle-num}
 \bibliography{homogenization}
\end{document}